\documentclass[cp1251%,draft % �����������������, ���� ����� �������� �����
               ]{jetp} % jetp.cls
\twocolumn % ����������������, ���� �� ����� �������������� �������
\begin{document}
{\English

\title{50 years in the Landau Institute environment}

\setaffiliation1{Low Temperature Laboratory, Aalto University,  \\ P.O. Box 15100, FI-00076 Aalto, Finland}

\setaffiliation2{Landau Institute for Theoretical Physics, \\  acad. Semyonov av., 1a, 142432,
Chernogolovka, Russia}

\setauthor{G.~E.}{Volovik}{12}
\email{volovik@boojum.hut.fi}
 
\rtitle{Landau Institute}
\rauthor{G.~E. Volovik}

\maketitle

%%%%%%%%%%%%%%%%%%%%%%%%%%%%%%%%%%%%%%%%%%%%%%%%%%%%%%%%%%%%%%%%%%%%%%%%

\section{Introduction}

 It is impossible to review all the fundamental works of Landau Institute, which made important contribution to physics in general.
Here I will discuss only a part of these works, only those which directly influence my own work during 50 years (1968-2018). Khalatnikov created the unique Institute, where practically all important areas of theoretical physics have been represented, opening  broad possibilities  for collaboration. The multdisciplinary  environment of the  Landau Institute was the main element of inspiration.

 \section{Khalatnikov lectures and emergent relativity in superfluids}
\label{KhalatLecture}

The Khalatnikov lectures on superfluid $^4$He for students in Kapitza Institute  led me to strange connection between superfluid hydrodynamics and special and general relativity (probably in 1967). 
The relativistic character of the flow of superfluid $^4$He is manifested at low temperature, where the normal component is represented by "relativistic" excitations with linear spectrum -- phonons. The Eq.(3.13) in the Khalatnikov book \cite{KhalatnikovBook} shows the free energy of phonons in the presence of the counterflow ${\bf w}={\bf v}_{\rm n}-{\bf v}_{\rm s}$ (the flow of normal component of the liquid with respect to the superfluid component):
\begin{equation}
F_{\rm ph}(T,w)=\frac {F_{\rm ph}(T) }{\left( 1 -w^2/c^2\right)^2} \sim 
\left(\frac{T}{\sqrt{1 -w^2/c^2}}\right)^4\,,
\label{BPhonons}
\end{equation}
with $F_{\rm ph}(T)  \propto T^4$.
This equation recalls the Tolman law in general relativity, $T({\bf r})= T /\sqrt{g_{00}({\bf r})}$, which may suggest (and indeed suggested for me)  the idea that the vacuum is superfluid; the matter is represented by excitations; the gravitational field is the result of the flow of the vacuum with superfluid velocity $v_{\rm s}^2/2= GM/r$ (in full thermodynamic equilibrium  one has ${\bf v}_{\rm n}=0$, while ${\bf v}_{\rm s}$ may depend on coordinates, and $g_{00}({\bf r})=1 -w^2/c^2=1 -v_{\rm s}^2({\bf r})/c^2$); and  GR is some extension of the Landau-Khalatnikov two-fluid hydrodynamics. My attempt to share this crazy idea with Sharvin, who governed the student seminar at the Kapitza Institute, was not successful. I got the response that GR is a very beautiful theory, and it should not be spoiled by unjustified models. Now I fully agree with his absolutely correct  response.

%%%%%%%%%%%%%%%%%%%%%%%%%%%%%%%%%%%%%%%%
%%%%%%%%%%%%%%%%%%%%%%%%%%%%%%%%%%%%%%%%
\begin{figure}[top]
\centerline{\includegraphics[width=\linewidth]{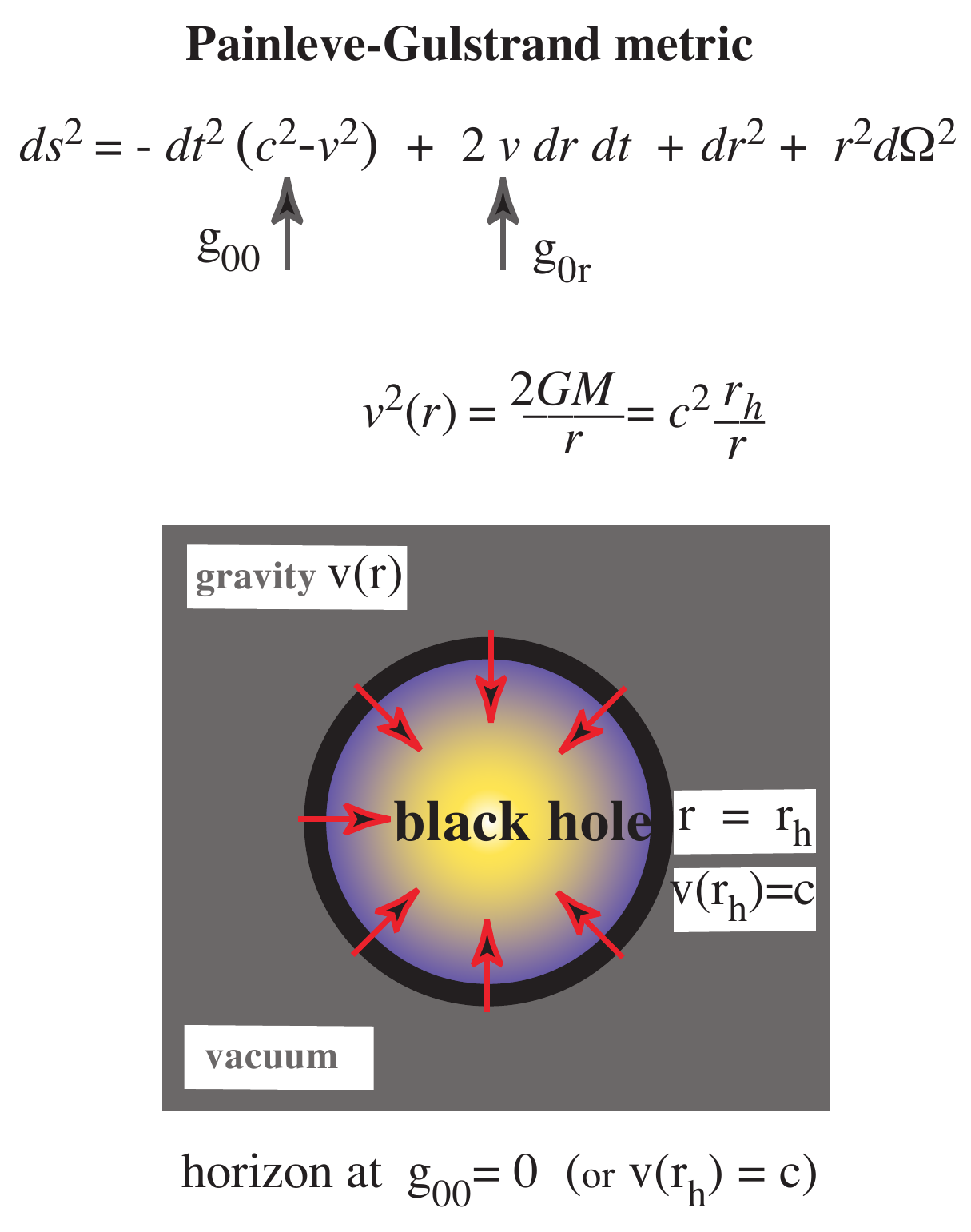}}
\label{BlackHole} 
  \caption{Black hole in Painlev\'e-Gullstrand metric.
}
\end{figure}
%%%%%%%%%%%%%%%%%%%%%%%%%%%%%%%%%%%%%%%%
%%%%%%%%%%%%%%%%%%%%%%%%%%%%%%%%%%%%%%%%

However, later on the papers by Unruh on black hole analogs in moving liquids appeared \cite{Unruh1981,SchutzholdUnruh}. The effective metric experienced by sound waves in liquids (or correspondingly phonons in superfluids)  became known as the acoustic metric, and the flow of the liquid with the acoustic horizon as the river model of black holes
\cite{RiverModel}. In GR, the flow metric with acoustic horizon corresponds to the  Painlev\'e-Gullstrand (PG) metric \cite{Painleve},
see Fig. 1. %\ref{BlackHole}. 
Following this stream with Ted Jacobson, we considered the Painlev\'e-Gullstrand metric, which emerges for the fermionic excitations in superfluid $^3$He -- the Bogoliubov-Nambu quasiparticles --  and discussed the possibility to create the analogs of black hole and white hole horizons using moving textures, solitons \cite{JacobsonVolovik1998}. 

The PG metric is useful for consideration of the processes inside and across the horizon such as Hawking radiation. 
Both for the real black hole and for  its condensed matter analog, the Hawking radiation can be considered as semiclassical quantum tunneling across the horizon, see Ref. \cite{Volovik1999} for the analog black hole and  
Ref. \cite{ParikhWilczek2000} for the real black hole. 

The acoustic metric allows us to simulate many different spacetimes.  I remember the talk by Polyakov in 1981, where he mentioned that the Minkowski signature in GR may emerge from the Euclidean one in a kind of the symmetry breaking phase transition. The effective Minkowski-to-Euclidean signature change can be probed in particular using  acoustic metric for Nambu-Goldstone mode in  the Bose-Einstein condensate of magnons \cite{NissinenVolovik2017,Autti2018}.

In 2003, many different analogies between the condensed matter on one side and relativistic quantum fields   and  gravity on the other side have been collected in the book \cite{Volovik2003}. In the fermionic  Weyl and Dirac materials, emergent gravity is formulated in terms of tetrad fields (see also the recent papers
\cite{Volovik2016c,NissinenVolovik2018}), instead of the metric gravity emerging in bosonic condensed matter systems. Moreover, in Weyl materials, gravity emerges together with all the ingredients of the  relativistic quantum field theories (relativistic spin, chiral fermions, gauge fields,
$\Gamma$-matrices,  etc.), see also Sec.\ref{GribovSec}.

 \section{Iordanskii and  macroscopic quantum tunneling}

In Landau Institute my supervisor was Iordanskii -- the author of thermal nucleation of vortices \cite{Iordanskii1965} and, together with Finkelshtein, of quantum formation of nucleation centers in a metastable crystal \cite{IordanskiiFinkelshtein1972,IordanskiiFinkelshtein1973}. Iordanskii suggested to me the problem of quantum nucleation of vortices in superfluids. This resulted in the paper on vortex nucleation in moving superfluids by quantum tunneling
\cite{Volovik1972}. 

The main difference from other types of macroscopic quantum tunneling is that the
role of the canonically conjugate quantum variables is played by the $z$ and $r$ coordinates of the vortex ring. This provides the volume law for the vortex instanton: the action contains the topological term $S_{\rm top}=2\pi \hbar n V_L=2\pi \hbar N_L$, where $n$ is the particle density; $V_L$ is the volume inside the surface swept by vortex line between its nucleation and annihilation; and $N_L$ is the number of atoms inside this volume, see Sec. 26.4.3 in the book \cite{Volovik2003}. For other linear topological defects and for fundamental  strings the  area law
is applied \cite{PolyakovString}.

The further development of this macroscopic quantum tunneling in superconductors can be found in the review paper \cite{Blatter1994}, where most authors are from Landau Institute:  Feigel'man,  Geshkenbein and Larkin.

 \section{Khalatnikov, superfluid $^3$He, paradox of angular momentum and chiral anomaly}

It is not surprizing that the epoch of superfluid $^3$He  in Landau Institute has been initiated by Khalatnikov. 
My participation in this programme started with collaboration with Khalatnikov and Mineev on the extension of the Landau-Khalatnikov hydrodynamics of superfluid $^4$He to  the dynamics of mixture of Bose and Fermi superfluids \cite{VolovikMineevKhalat1975}. The most interesting topic there was the  Andreev-Bashkin effect, when the superfluid current of one component depends also on the superfluid velocity of the other component  \cite{AndreevBashkin1975}. The seminal paper by Andreev and Bashkin  \cite{AndreevBashkin1975} had been published just in the previous issue of JETP, which demonstrated the traditionally close connection between the Landau and Kapitza Institutes.

The first attempts to extend the Landau-Khalatnikov hydrodynamics  to the hydrodynamics of the chiral superfluid $^3$He-A immediately showed some strange paradox related to the intrinsic angular momentum
of the liquid with  Cooper pairing into the $p+ip$ state \cite{Volovik1975,VolovikMineevHydro1976}. The calculated magnitude of the 
dynamical angular momentum was by factor $(\Delta_0/E_F)^2$ smaller than the expected angular momentum of the stationary state, $L_z=\hbar N/2$, which corresponds to $\hbar$ for each of $N/2$ Cooper pairs,  where $N$ is the number of atoms. With $\Delta_0$ being the gap amplitude in the fermionic quasiparticle spectrum and $E_F$ the Fermi energy, this  factor is very small, $(\Delta_0/E_F)^2  \sim 10^{-5}$.

Only about ten years later some understanding was achieved \cite{BalVolKon1986,Volovik1986a}
that the source of the angular momentum paradox and of the other related paradoxes in the $^3$He-A dynamics was the the analog of the chiral anomaly in RQFT, see Eq.(\ref{Anomaly}). The chiral anomaly is realized in the quantum vacuum with Weyl fermions, and the topologically protected Weyl fermions emerge in the chiral superfluid $^3$He-A \cite{Volovik2003}.
The Khalatnikov-Lebedev hydrodynamics of chiral supefluid $^3$He \cite{KhalatnikovLebedev1977} has to be  modified to include the chiral anomaly effects. 

The effect of chiral anomaly has been experimentally verified in dynamics of skyrmions in $^3$He-A \cite{Bevan1997}, see Sec. \ref{KopninIordanskii}.
See also the recent discussion on the connection of chiral anomaly to the angular momentum paradox in chiral superfluids and superconductors \cite{Volovik2014,Tada2018}.

 \section{Polyakov monopole and vortex with free end}

%%%%%%%%%%%%%%%%%%%%%%%%%%%%%%%%%%%%%%%%
%%%%%%%%%%%%%%%%%%%%%%%%%%%%%%%%%%%%%%%%
\begin{figure}[top]
\centerline{\includegraphics[width=0.8\linewidth]{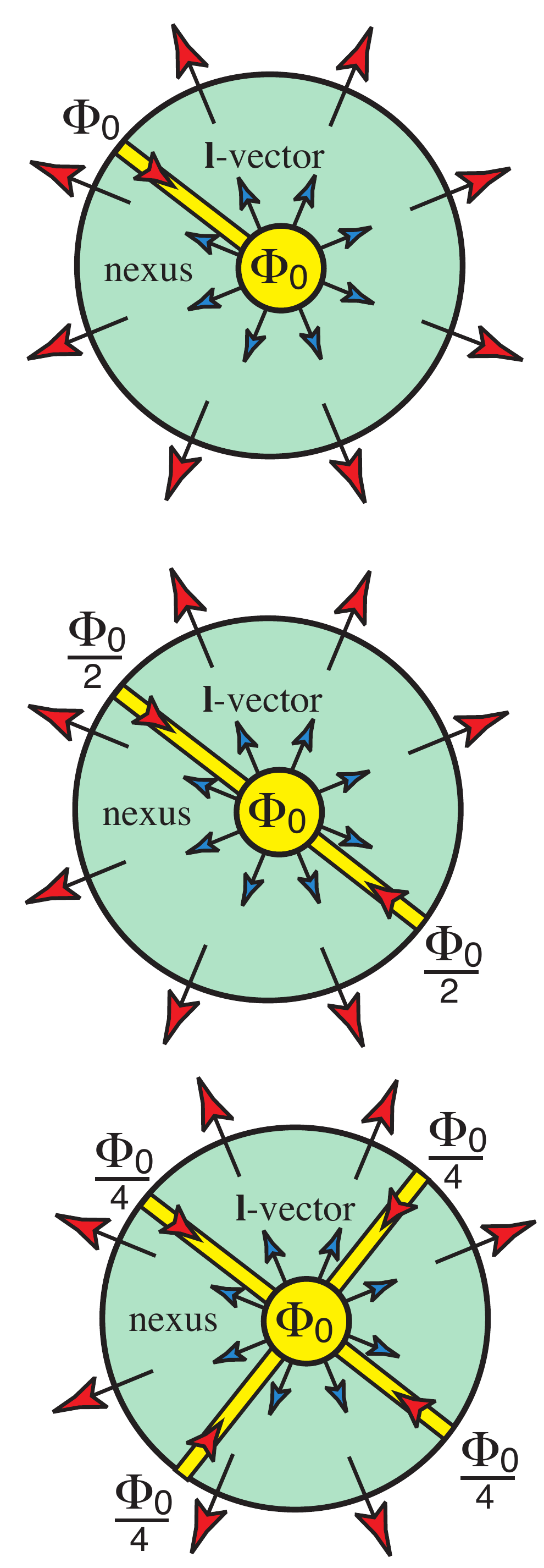}}
\label{four} 
  \caption{Magnetic monopole in chiral superconductor -- the analog of the Nambu monopole. ({\it Top}): monopole, which terminates the doubly quantized vortex, $N=2$. ({\it Middle}): the same monopole terminates two vortices with $N=1$. ({\it Bottom}): nexus -- monopole with four half-quantum ($N=1/2$)  vortices  -- the Alice strings in Fig. 4.%\ref{AliceString}. 
 Red arrows show direction of magnetic flux, which is brought to the monopole by vortices, and then radially propagates from the monopole. The blue arrows is the field of the orbital vector ${\bf l}$, which forms the hedgehog.   }
\end{figure}
%%%%%%%%%%%%%%%%%%%%%%%%%%%%%%%%%%%%%%%%
%%%%%%%%%%%%%%%%%%%%%%%%%%%%%%%%%%%%%%%%

Polyakov gave a talk at Landau Institute seminar in Chernogolovka on the hedgehog-monopole, which later got the name 't Hooft-Polyakov monopole \cite{Polyakov1974,tHooft1974}.
Inspired by this talk, we with Mineev suggested the analog of magnetic monopole in $^3$He-A \cite{VolovikMineev1976a} (the same suggestion was made by Blaha  \cite{Blaha1976}). As distinct from the  't Hooft-Polyakov monopole, this monopole terminates the linear defects -- vortices and strings. It can terminate either  singular doubly quantized vortex with $N=2$ in Fig. 2 %\ref{four}.  
({\it top}), which we called the vortex with free end; or two singly quantized vortices with $N=1$ in Fig. 2 %\ref{four}. 
({\it middle}); or four half-quantum vortices with $N=1/2$ in  Fig. 2
%\ref{four}.  
({\it bottom}).

In electroweak theory such  monopole terminating the electroweak string  is known as the Nambu monopole \cite{Nambu1977}.

The condensed matter analog of magnetic monopole, which terminates the string, has been been observed in cold gases \cite{Mottonen2014}.

\section{Ferromagnetic hedgehog as magnetic monopole in synthetic field}

The monopole-hedgehog topic started by Polyakov and also some vague ideas on the possible emergence of gauge fields had the following development. It appeared that  in ferromagnets,  the Berry phase  gives rise to a synthetic electromagnetic field \cite{Volovik1987c}:
\begin{equation}
F_{ik}=\partial_i A_k -\partial_k A_i = - \frac{1}{2} {\bf m} \cdot(\partial_ i{\bf m} \times \partial_k{\bf m} )\,,
\label{Bfield}
\end{equation} 
\begin{equation}
E_{i}=-\partial_t A_i -\partial_i A_0 = \frac{1}{2} {\bf m} \cdot(\partial_ t{\bf m} \times \partial_i{\bf m} )\,,
\label{Efield}
\end{equation} 
where ${\bf m}$ is the unit vector of magnetization.
The effective electric and magnetic fields are physical: they act on electrons in ferromagnets and produce 
in particular the spin-motive force, induced by a time and spatial derivatives of magnetization \cite{Volovik1987c,Barnes2007}. This force is proportional to  ${\bf E}-{\bf v}\times {\bf B}$ similar to that in quantum electrodynamics. The spin-motive force is enhanced in the presence of Dzyaloshinskii-Moriya 
interaction \cite{Yamane2018}. 

The hedgehog in ferromagnets in Fig. 3 %\ref{monopoles} 
({\it left}) appeared to be
 the monopole in the Berry phase magnetic field ${\bf B}$ \cite{Volovik1987c}. This somewhat reminds the Polyakov hedgehog in the Higgs field, which at the same time represents the magnetic monopole.
However, as distinct from Weyl material scenario of the emergent gauge field, in this Berry phase scenario  the full analogy with the electromagnetic field is missing.

%%%%%%%%%%%%%%%%%%%%%%%%%%%%%%%%%%%%%%%%
%%%%%%%%%%%%%%%%%%%%%%%%%%%%%%%%%%%%%%%%
\begin{figure}[top]
\centerline{\includegraphics[width=\linewidth]{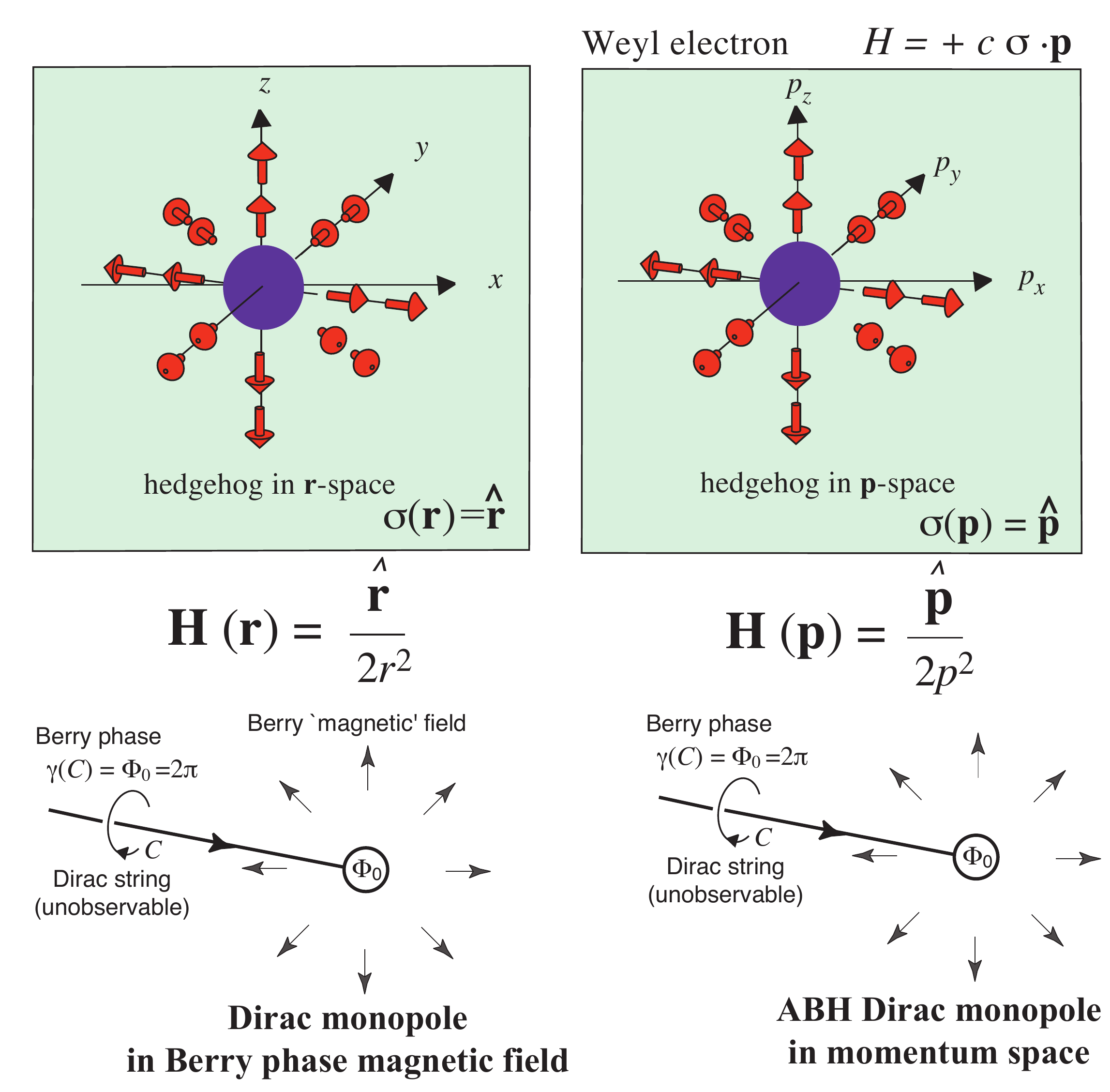}}
\label{monopoles} 
  \caption{Berry phase magnetic monopoles in real and momentum space. ({\it left}): Berry phase magnetic monopole  in ferromagnets. ({\it right}): Abrikosov-Beneslavskii-Herring monopole in momentum space, which is responsible for the topological stability of Weyl fermions. Both have unobservable Dirac string with $2\pi$ winding of the Berry phase.
 }
\end{figure}
%%%%%%%%%%%%%%%%%%%%%%%%%%%%%%%%%%%%%%%%
%%%%%%%%%%%%%%%%%%%%%%%%%%%%%%%%%%%%%%%%

\section{Novikov, topology, Alice string,  Berezinzkii}

 During his seminar talk at the Landau Institute on the hedgehog-monopole,  Polyakov mentioned that mathematicians claim that the hedgehog cannot be destroyed for topological reasons. This led to the intensive study of topology in physics and discussions with the members of the Novikov group in Landau Institute (Bogoyavlensky, Grinevich, and others).
Also the Anisimov-Dzyaloshinskii paper on disclinations appeared \cite{AnisimovDzyaloshinskii1973}, where the variety of structures in liquid crystals was discussed.
 With Mineev we wanted to understand how and why these and other structures including our vortex with free end (the analog of Nambu monopole) were topologically stable  or not.

%%%%%%%%%%%%%%%%%%%%%%%%%%%%%%%%%%%%%%%%
%%%%%%%%%%%%%%%%%%%%%%%%%%%%%%%%%%%%%%%%
\begin{figure}[top]
\centerline{\includegraphics[width=\linewidth]{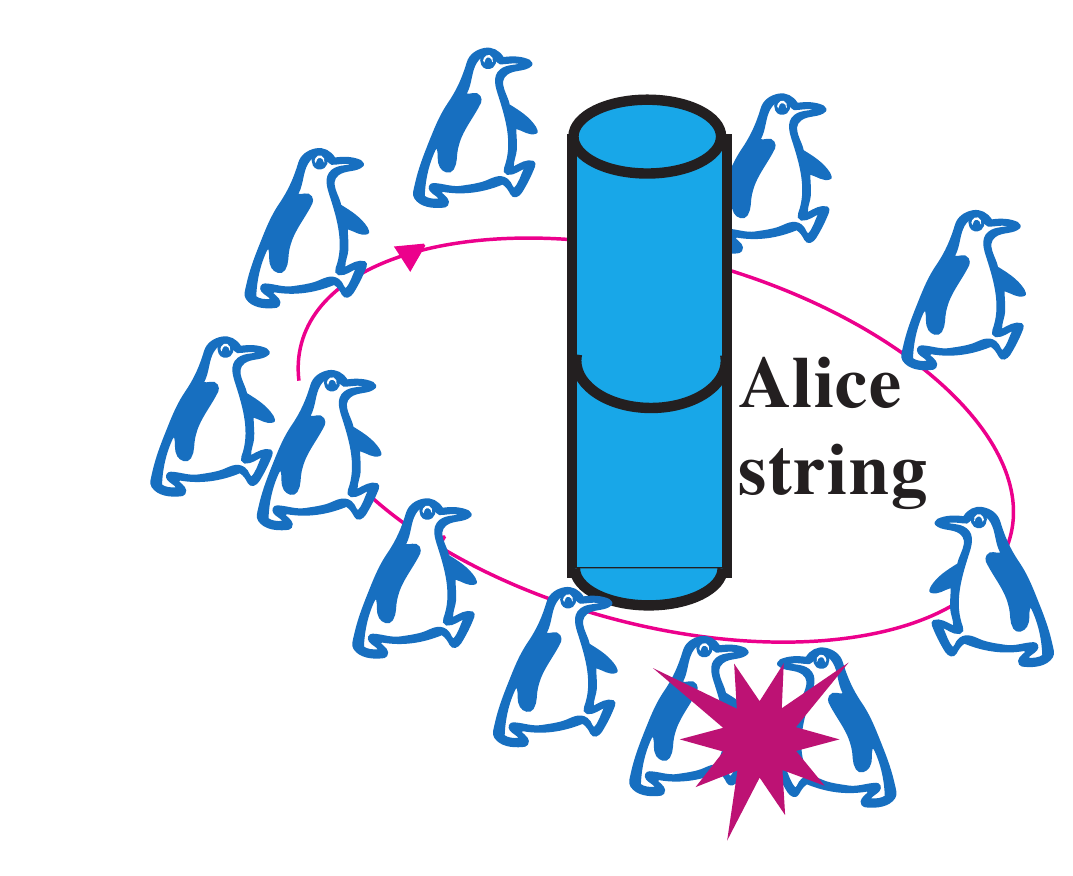}}
\label{AliceString} 
  \caption{Half-quantum vortex as Alice string \cite{Schwarz1982}.
Matter continuously transforms to antimatter after circling Alice string.
Two penguins start to move in opposite directions around the string. When they meet each other, they annihilate.
}
\end{figure}
%%%%%%%%%%%%%%%%%%%%%%%%%%%%%%%%%%%%%%%%
%%%%%%%%%%%%%%%%%%%%%%%%%%%%%%%%%%%%%%%%

This led us to the homotopy group classification of topological structures \cite{VolovikMineev1976b,VolovikMineev1977e}. Among these structures, some unexpectred exotic topological objects were suggested, such as the half-quantum vortex in $^3$He-A \cite{VolovikMineev1976b}.
In RQFT, the analog of the half-quantum vortex is the Alice string in Fig. 4
%\ref{AliceString} 
discussed by Schwarz \cite{Schwarz1982},
who collaborated with Belavin and Polyakov in the instanton problem \cite{Schwarz1975}.
Experimentally the half-quantum vortices were observed only 40 years later, first in the time-reversal symmetric polar phase of $^3$He \cite{Autti2016}, and finally in the chiral $^3$He-A \cite{Makinen2018}. 
Moreover, it was found that they survive the phase transition to $^3$He-B, where the half-quantum vortex is topologically unstable: it becomes the termination line of the non-topological domain wall - the analog of Kibble cosmic wall \cite{Kibble1982}.

%%%%%%%%%%%%%%%%%%%%%%%%%%%%%%%%%%%%%%%%%%%%%%%%%%%%%%%%%
%%%%%%%%%%%%%%%%%%%%%%%%%%%%%%%%%%%%%%%%%%%%%%%%%%%%%%%%%
\begin{figure}[t]
\includegraphics[width=\linewidth]{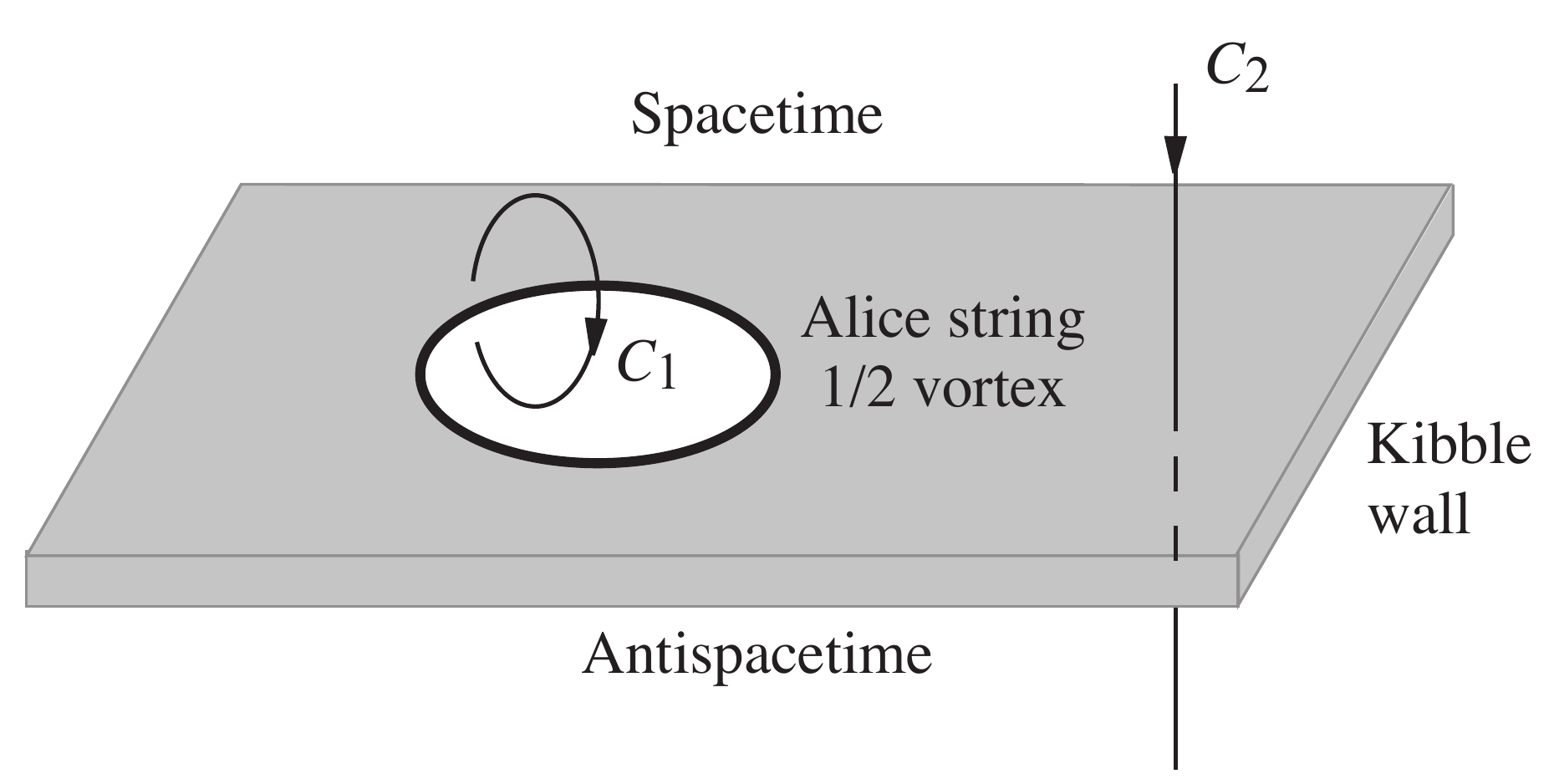}
\caption{  In $^3$He-B,  Alice string (the half-quantum vortex) becomes the termination line of non-topological domain wall -- the Kibble wall \cite{Makinen2018}.  There are two roads to antispacetime: the safe route around the Alice string (along the contour $C_1$) or dangerous route along $C_2$ across the Kibble wall\cite{Volovik2019}.
}
\label{TwoRoads_Fig}
\end{figure}
%%%%%%%%%%%%%%%%%%%%%%%%%%%%%%%%%%%%%%%%%%%%%%%%%%%%%%%%%
%%%%%%%%%%%%%%%%%%%%%%%%%%%%%%%%%%%%%%%%%%%%%%%%%%%%%%%%%

Vortices in $^3$He-A are described by the $Z_4$ homotopy group, which means that  $1/2 +1/2+1/2+1/2=1+1=2=0$, and thus 4 half-quantum vortices can terminate at the monopole in Fig. 2. %\ref{four}.  
Also, in systems such as liquid crystals, the topological defects -- disclinations -- may obey even the non-Abelian homotopy groups. All this caused the interest from Berezinzkii to the possibility of the extension of the BKT transition \cite{Berezinskii1971,KT1973} to the more general symmetry breaking patterns, but unfortunately he passed away in 1980.

Novikov himself also participated in the $^3$He business. In particular, he resolved the paradox related to the number of the Nambu-Goldstone (NG) modes in $^3$He-A: in the weak coupling limit there are 9 NG modes, but only 8  broken symmetry generators  \cite{VolovikKhazan1982}. Novikov formulated the new counting rule \cite{Novikov1982}: the number of NG modes coincides with the dimension of the tangent space. The mismatch between the total number of NG bosons and the number of broken symmetry generators equals the number of extra flat directions in the Higgs potential.

\section{Novikov, topology,  skyrmions}

The next our step in the classification of topological structures in condensed matter was triggered again by the Landau Institute environment: the
Belavin-Polyakov topological object in 2D Heisenberg ferromagnets \cite{BelavinPolyakov1975}, dynamical solitons discussed by the Zakharov group \cite{Solitons1986}, and discussions with members of the Novikov group (Golo and Monastyrsky \cite{GoloMonastyrsky1978}). All this led us to classification of the continuous structures in terms of  relative homotopy groups \cite{MineyevVolovik1978}. 

These structures include in particular  the analogs of the 3D skyrmions: the particle-like solitons described by the $\pi_3$ homotopy group \cite{VolovikMineev1977b} (which got the name Shankar monopole \cite{Shankar1977}). The isolated 3D skyrmions have been observed in cold gases \cite{Mottonen2018}. In superfluid $^3$He, it is still difficult to stabilize the isolated skyrmions, but the $\pi_3$ objects at the crossing of 1D and 2D topological solitons have been created experimentally \cite{Ruutu1994}, see the topological analysis of these combined objects by Makhlin and Misirpashaev \cite{Makhlin1995}.

 \section{Kopnin and vortex  skyrmion lattice}

%%%%%%%%%%%%%%%%%%%%%%%%%%%%%%%%%%%%%%%%
%%%%%%%%%%%%%%%%%%%%%%%%%%%%%%%%%%%%%%%%
\begin{figure}[top]
\centerline{\includegraphics[width=\linewidth]{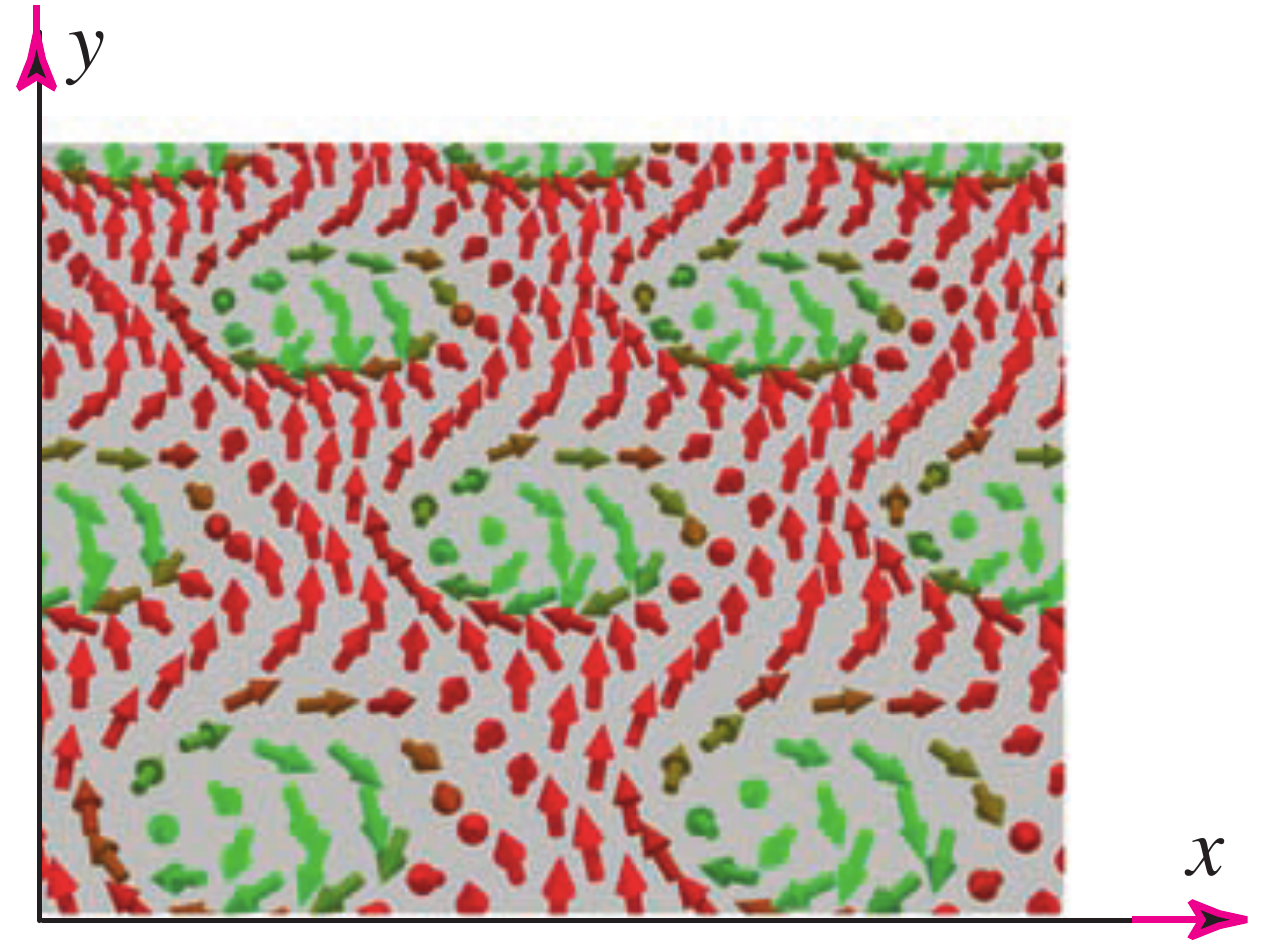}}
\label{SkyrmionFig} 
  \caption{Skyrmion lattice in rotating chiral superfluid. Each cell has two quanta of circulation of superfluid velocity.
}
\end{figure}
%%%%%%%%%%%%%%%%%%%%%%%%%%%%%%%%%%%%%%%%
%%%%%%%%%%%%%%%%%%%%%%%%%%%%%%%%%%%%%%%%

As distinct form 3D skyrmions, the 2D skyrmions are typical in the Helsinki experiments with the chiral $^3$He-A in a rotating cryostat.
 In superfluid $^3$He-A, the vorticity can be continuous (nonsingular) and can form a periodic texture in the rotating vessel -- the 2D skyrmion lattice, which we discussed with Kopnin \cite{VolovikKopnin1977}, see Fig. 6.%\ref{SkyrmionFig}. 
This paper opened the collaboration with Kopnin.
The 2D skyrmions have been later identified in NMR experiments
\cite{Seppala1984}. 
Subsequently, the change of the topological charge of skyrmion was observed in ultrasound experiments\cite{Pekola1990}.    Sweeping the magnetic field, we could see the first order topological transition between different configurations. In small fields, the skyrmion has nontrivial charges both in the orbital and in the spin vector fields, $N_l=1$ and $N_d=1$. In high field, the skyrmion looses one of the winding numbers, $N_l=1$ and $N_d=0$. 

The similar skyrmion lattice has been suggested by Kopnin for  anisotropic superconductivity, in which the symmetry breaking pattern is
$(SU(2)_{S}\times  U(1)_N)/Z_2 \rightarrow U(1)_{S_z-N/2}$
\cite{BurlachkovKopnin1987}.

 \section{Bekarevich-Khalatnikov theory of rotating superfluid}

%%%%%%%%%%%%%%%%%%%%%%%%%%%%%%%%%%%%%%%%
%%%%%%%%%%%%%%%%%%%%%%%%%%%%%%%%%%%%%%%%
\begin{figure}[top]
\centerline{\includegraphics[width=\linewidth]{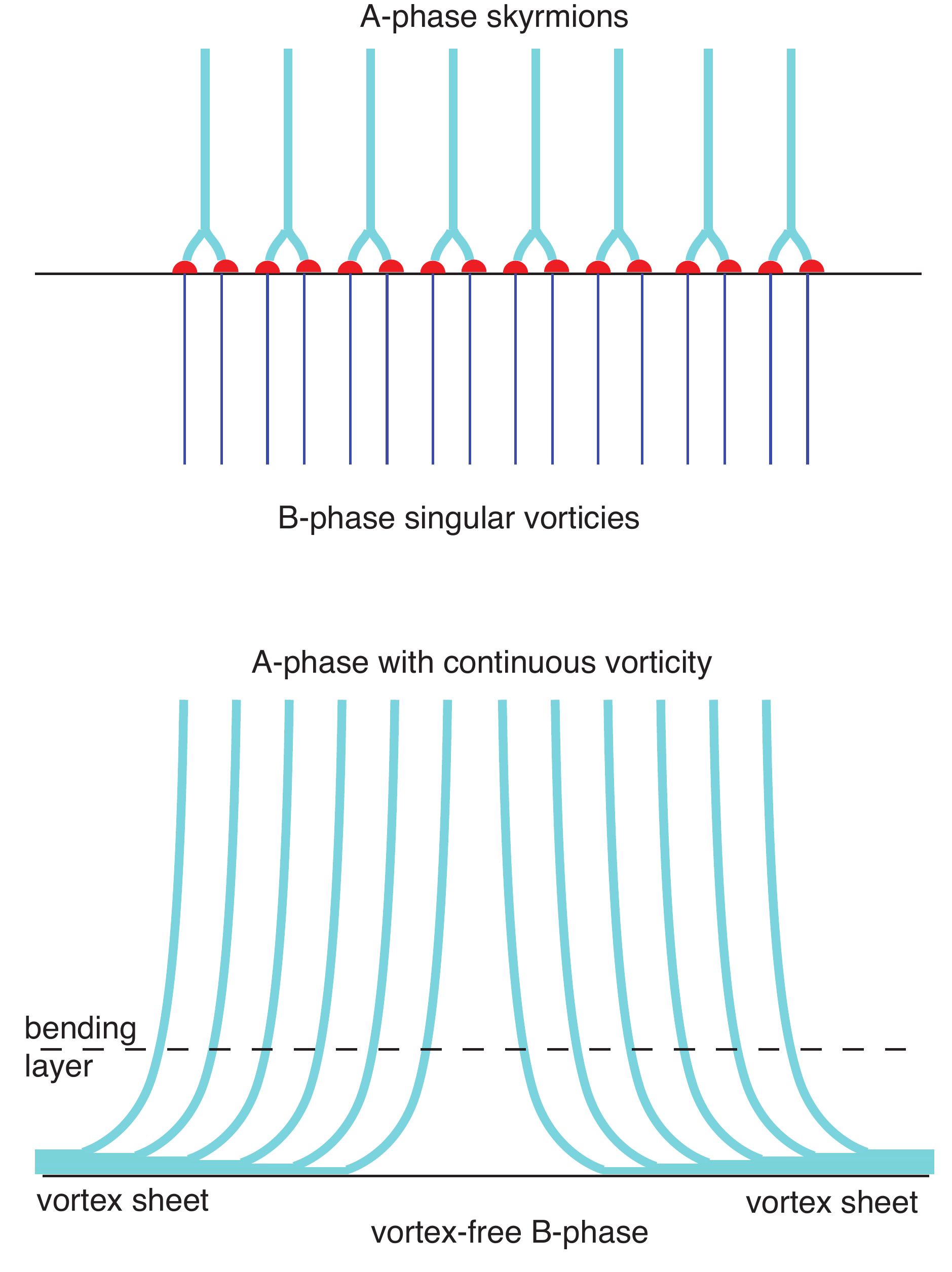}}
\label{SkyrmionInterface} 
  \caption{The interface between chiral superfluid $^3$He-A and non-chiral $^3$He-B in rotating cryostat. ({\it top}): Skyrmion lattice in A-phase transforms to the conventional vortex lattice in the B-phase. 
The end point of the B-phase vortex is the analog of Nambu monopole. ({\it bottom}): Skyrmion lattice transforms to the vortex sheet at the interface, while the B-phase is made vortex free. The bending of vorticity at the interface obeys the Bekarevich-Khalatnikov theory of rotating superfluids \cite{BekarevichKhalatnikov1961}.
}
\end{figure}
%%%%%%%%%%%%%%%%%%%%%%%%%%%%%%%%%%%%%%%%
%%%%%%%%%%%%%%%%%%%%%%%%%%%%%%%%%%%%%%%%

Fig. 7
%\ref{SkyrmionInterface}  
demonstrates the skyrmion lattice in rotating cryostat in the presence of the interface between $^3$He-A and $^3$He-B \cite{Hanninen2003}. Experimentally one can produce different pattern of rotating superfluids. In particular, one of the superfluids, the A-phase, contains the equilibrium number of vortices, while another one, the B-phase, is vortex-free, see Fig. 7 
%\ref{SkyrmionInterface} 
 ({\it bottom}). In this case  the vortices bend and form the vortex sheet. To describe this bending we used the hydrodynamic equations derived by 
Bekarevich and Khalatnikov \cite{BekarevichKhalatnikov1961}.

 \section{Interface instability: Korshunov, Kuznetsov, Lushnikov}
\label{InterfaceInstabilitySec}

%%%%%%%%%%%%%%%%%%%%%%%%%%%%%%%%%%%%%%%%
%%%%%%%%%%%%%%%%%%%%%%%%%%%%%%%%%%%%%%%%
\begin{figure}[top]
\centerline{\includegraphics[width=\linewidth]{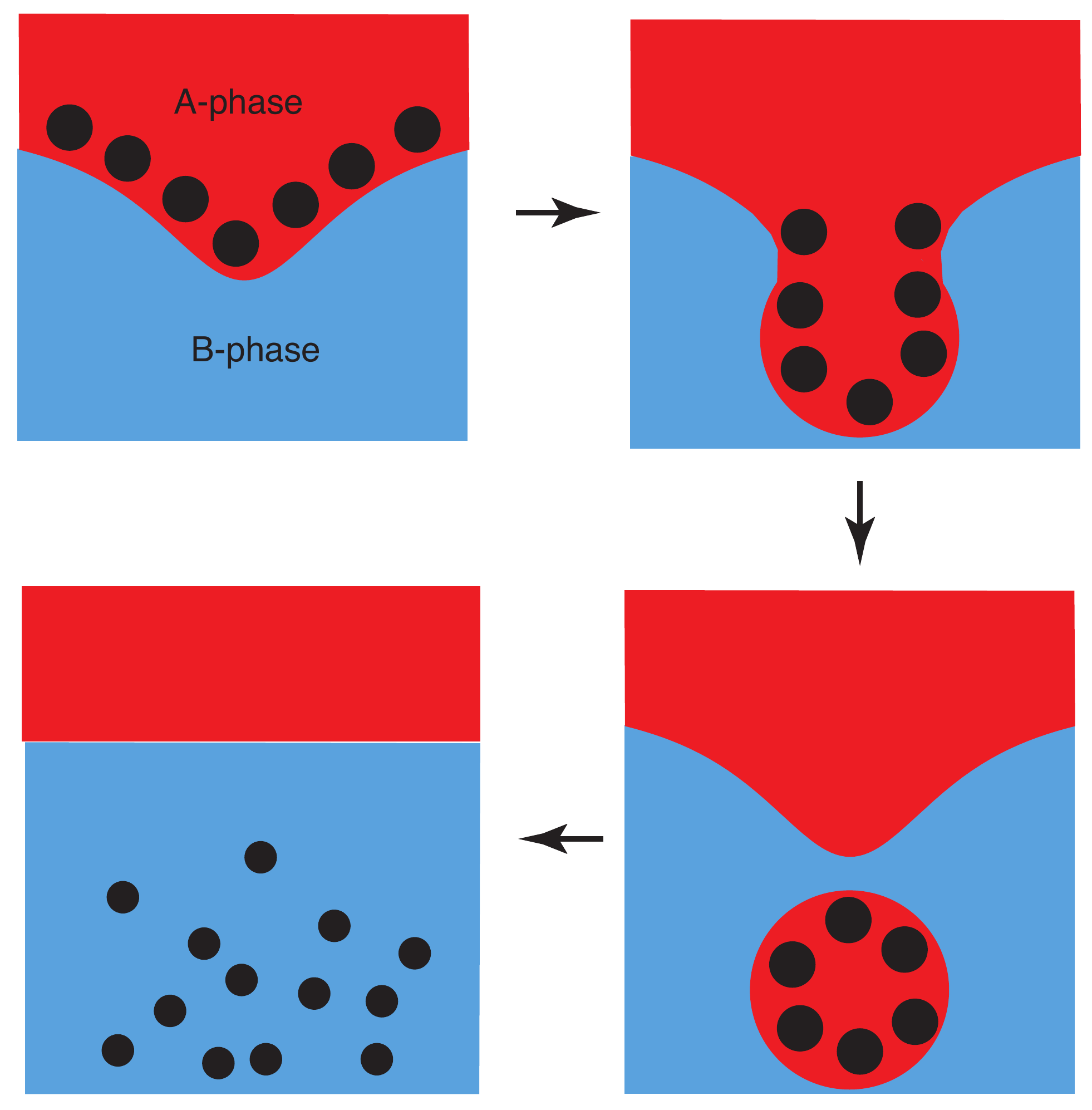}}
\label{KH} 
  \caption{Development of the Kelvin-Helmholtz type of instability at the interface between the vortex-full chiral superfluid $^3$He-A and vortex-free non-chiral $^3$He-B in rotating cryostat in Fig. 7. %\ref{SkyrmionInterface}.
 The instability leads to creation of vortices in $^3$He-B. The possible scenario: the droplet of the A-phase with vorticity concentrated in the skyrmions penetrates the AB interface, where vorticity transforms to the singular vortices. The NMR experiments show that the number of the B-phase vortices formed after instability is consistent with the wavelength of the critical ripplon.
}
\end{figure}
%%%%%%%%%%%%%%%%%%%%%%%%%%%%%%%%%%%%%%%%
%%%%%%%%%%%%%%%%%%%%%%%%%%%%%%%%%%%%%%%%

The arrangement in Fig. 7 %\ref{SkyrmionInterface} 
 allowed us to study experimentally the analog of the Kelvin-Helmholtz instability in superluids \cite{Blaauwgeers2002}. 
At some critical velocity of rotation the interface becomes unstable towards formation of ripplons at the interface \cite{Volovik2002}. Originally the Kelvin-Helmholtz (KH) instability in superfluids was studied by 
Korshunov \cite{Korshunov1991,Korshunov2002}.  Instead of the conventional KH instability of the interface between two fluids, Korshunov considered rather unusual case: using the Landau-Khalatnikov two-fluid model he studied the instability of the surface of the liquid under counterflow of the superfluid and normal components of the same liquid.  It happens that arrangement in the Fig. 7
%\ref{SkyrmionInterface} 
 is very similar to the Korshunov case. On one side of the interface, the vortex-full A-phase rotates together with the container, 
${<\bf v}_{\rm sA}>= {\bf v}_{\rm nA} =\mbox{\boldmath$\Omega$}\times {\bf r}$. On the other side of the interface, in the vortex-free B-phase, only the normal component rotates with container, while its superfluid velocity is at rest: ${\bf v}_{\rm sB}=0$, ${\bf v}_{\rm nB} =\mbox{\boldmath$\Omega$}\times {\bf r}$.
That is why the instability occurs due to the counterflow on the B-phase side, ${\bf w}_{\rm B}
={\bf v}_{\rm nB}-{\bf v}_{\rm sB}$.

The nonlinear stage of the KH instability has been considered by Kuznetsov and Lushnikov
\cite{KuznetsovLushnikov1995,Lushnikov2018}.
In our experiments, the development of the interface instability leads to penetration of the A-phase skyrmions through the interface to the B-phase, where they are finally transformed to the singular vortices. Fig. 8
%\ref{KH} 
demonstrates the possible scenario of this transformation. However, the complete analysis of the this process is still missing.

 \section{Vortex creation in a micro Big-Bang, Kopnin, Kamensky, Manakov }
\label{KibbleZurekSec}

%%%%%%%%%%%%%%%%%%%%%%%%%%%%%%%%%%%%%%%%
%%%%%%%%%%%%%%%%%%%%%%%%%%%%%%%%%%%%%%%%
\begin{figure}[top]
\centerline{\includegraphics[width=\linewidth]{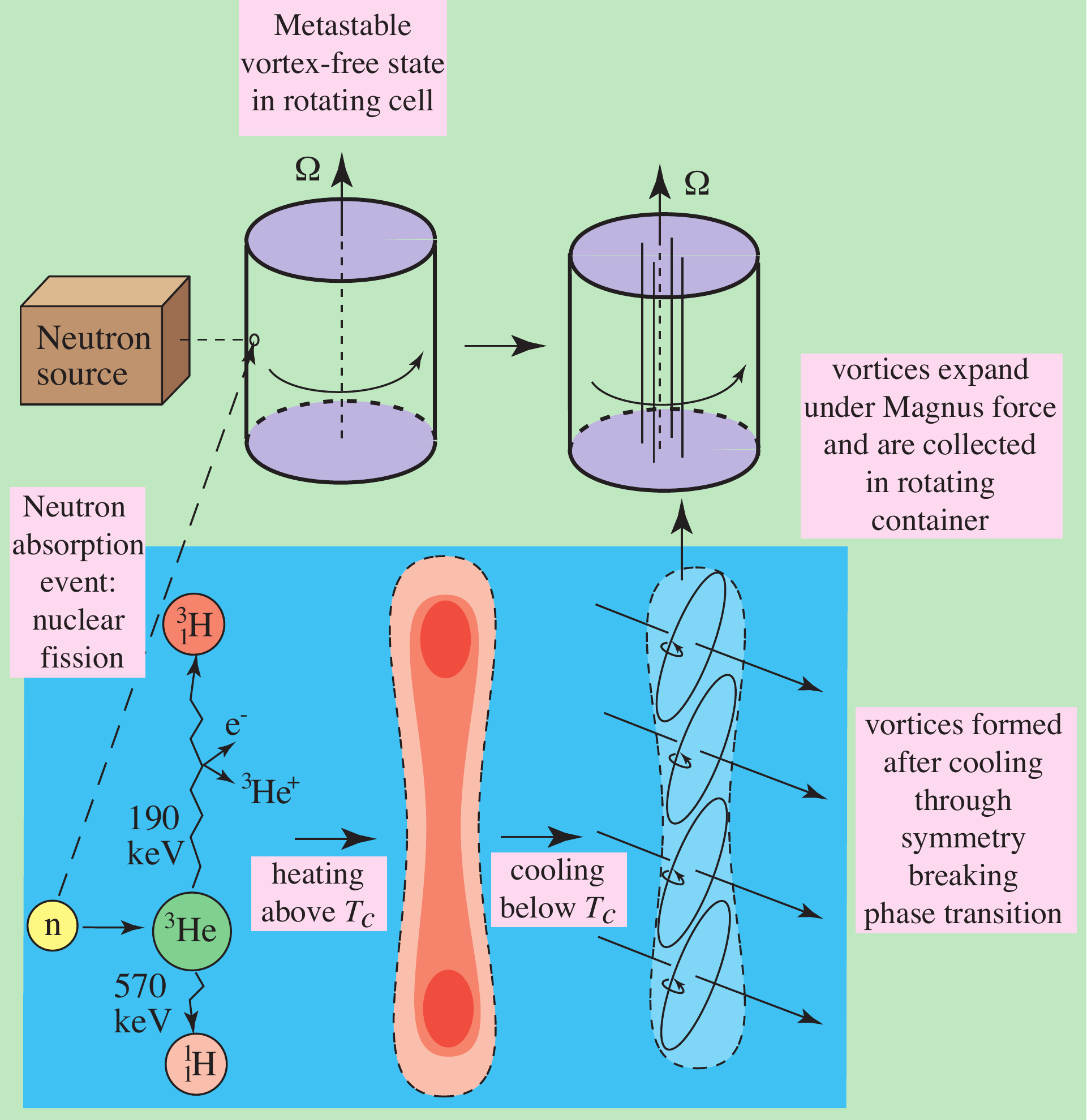}}
\label{BigBangFig} 
  \caption{Vortex formation in a micro Big-Bang event caused by neutron
irradiation \cite{Makhlin1996}.
}
\end{figure}
%%%%%%%%%%%%%%%%%%%%%%%%%%%%%%%%%%%%%%%%
%%%%%%%%%%%%%%%%%%%%%%%%%%%%%%%%%%%%%%%%

Another nonlinear out-of-equilibrium phenomenon studied experimentally in superfluid $^3$He had its origine  in the interplay of high-energy physics and cosmology. This is the nucleation of topological defects during the phase transition in expanding universe \cite{Kibble1976}, which got the name Kibble-Zurek mechanism of defect formation. In $^3$He-B, the Big-Bang event was simulated by neutron
irradiation, which caused nuclear reaction and heating of the bubble of about 100 $\mu m$ size above the transition temperature \cite{Makhlin1996}, see Fig. 9.
%\ref{BigBangFig}. 
Then the cooling of the bubble back through the second order phase transition to the broken symmetry state resulted in the creation of vortices measured in the NMR experiments. 

The explanation of the observed vortex creation in the frame of the  Kibble-Zurek scenario looks reasonable. Moreover, it is supported by the correct power-law dependence of the number of the created vortices on the velocity of the superfluid. Nevertheless the modfications and extensions of the Kibble-Zurek scenario were necessary in order to take into consideration the inhomogeneity of the process. In particular, the effect of propagation of the transition front was considered in our paper with Kibble \cite{KibbleVolovik1997} and in papers by Kopnin and co-authors \cite{KopninThuneberg1999,AransonKopninVinokur1999-2011}. But the original idea that vortices can be created by the propagating front of the second order phase transition belongs to Kamensky and Manakov \cite{ManakovKamensky1990}.

 \section{Instantons, BPST, chiral superfluids, superconductors}

%%%%%%%%%%%%%%%%%%%%%%%%%%%%%%%%%%%%%%%%
%%%%%%%%%%%%%%%%%%%%%%%%%%%%%%%%%%%%%%%%
\begin{figure}[top]
\centerline{\includegraphics[width=\linewidth]{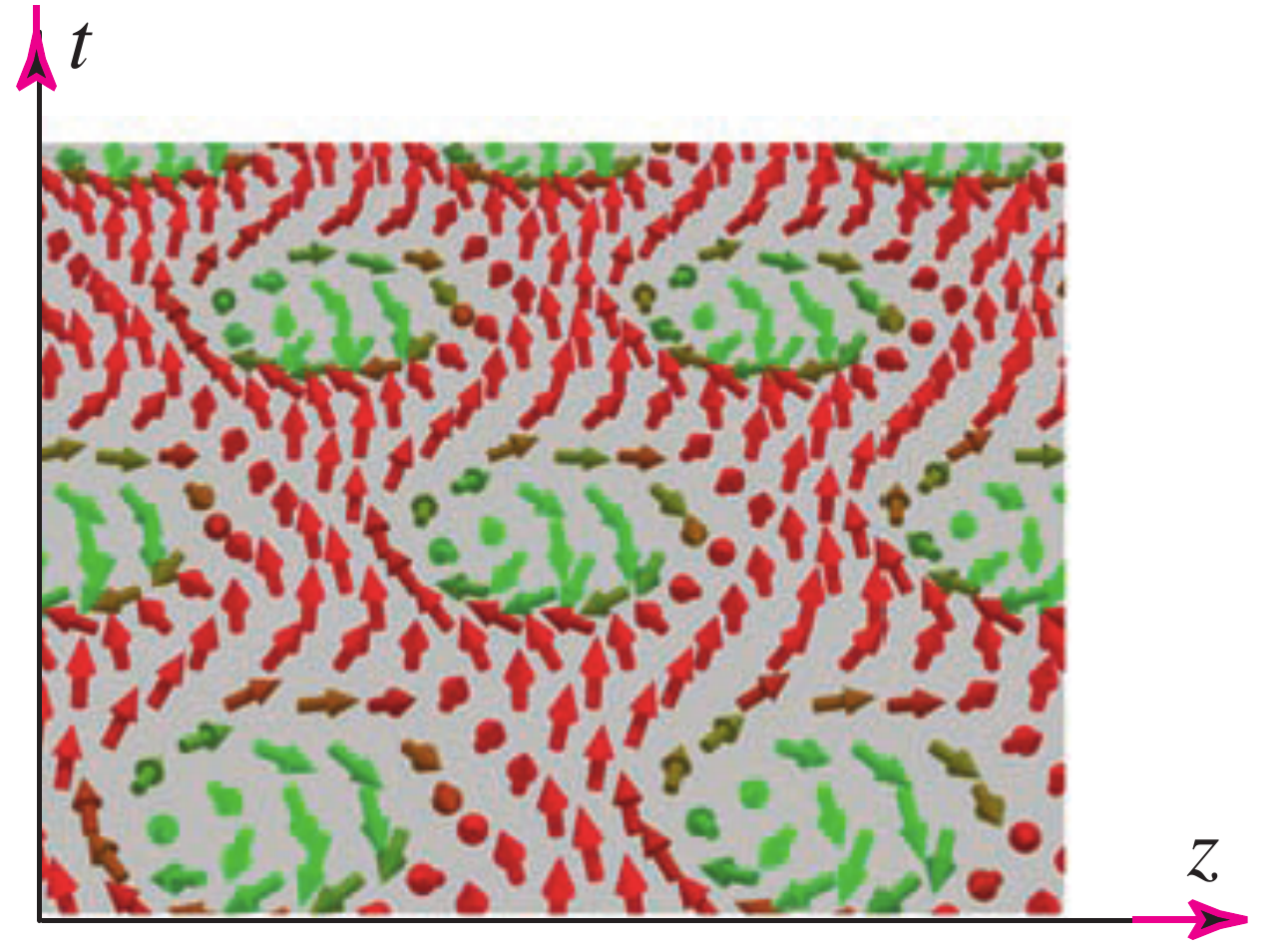}}
\label{InstantonFig} 
  \caption{Instanton lattice in dynamics of chiral superfluid. It looks like skyrmion lattice in Fig. 6, but in the $(z,t)$-plane.
}
\end{figure}
%%%%%%%%%%%%%%%%%%%%%%%%%%%%%%%%%%%%%%%%
%%%%%%%%%%%%%%%%%%%%%%%%%%%%%%%%%%%%%%%%

The  Poyakov \cite{Polyakov1975} and the Belavin-Polyakov-Schwartz-Tyupkin (BPST)  \cite{Schwarz1975}  instantons inspired the study of the instanton structures in condensed matter.  The $1+1$ instanton lattice \cite{Volovik1978} in Fig. 10
%\ref{InstantonFig} 
served for explanations of the  oscillations observed in the counterflow experiments in chiral superfluid $^3$He-A \cite{Krusius1976}. This $1+1$ lattice in the the $(z,t)$ plane is the counterpart of the $2+0$ skyrmion lattice in Fig. 6, where $z$ is the coordinate along the counterflow.
Similar $1+1$ instanton structure, but in terms of of the $(z,t)$ counterparts of Abrikosov vortex lattice,  is discussed for superconductors by Ivlev and Kopnin \cite{IvlevKopnin1978}.

\section{Dzyaloshinskii, spin glasses, general hydrodynamics and dimensionless physics}

The natural objects for studies of the topologically stable structures are magnetic materials, where the main expert in Landau Institute was Dzyaloshinskii -- the author of Dzyaloshinskii-Moriya interaction
\cite{Dzyaloshinskii1958,Moriya1960}. The common interest in topological defects in magnetic materials led to our collaboration. We considered frustrations in spin glasses introduced by Villain \cite{Villain1977} and suggested that on the macroscopic hydrodynamic level, the frustrations can be described in terms of the topological defects -- disclinations, which destroy the long range magnetic order \cite{DzyalVol1978}.  The continuous distribution of disclinations and their dynamics can be described using the effective gauge fields: the $U(1)$ gauge field in XY spin glasses, and $SU(2)$ gauge field in the Heisenberg spin glasses.
This provides another scenario for emergent gauge fields in addition to the Berry phase scenario in ferromagnets and Weyl point scenario in $^3$He-A and in Weyl semimetals.

The  hydrodynamics of systems with distributed defects was then extended to superfluids with vortices, and crystals with dislocations and disclinations \cite{DzyalVol1980,VolovikDotsenko1979}. The  relevant gauge fields which describe the distributed dislocations and disclinations are correspondingly the torsion and Riemann curvature in the formalism of general relativity.

It is important that the elasticity tetrads $E^{~a}_\mu(x)$, which describe elastic deformations of the crystal lattice, are expressed in terms of a system of deformed crystallographic coordinate planes, surfaces of constant phase $X^a(x)=2\pi n^a$ \cite{DzyalVol1978,NissinenVolovik2018b,NissinenVolovik2018c}. 
The tetrads $E^{~a}_\mu(x)= \partial_\mu X^a(x)$ have dimension of inverse length (or inverse time).
  When these elasticity tetrads are applied to general relativity (the so called superplastic vacuum \cite{KlinkhamerVolovik2019}), one obtains that 
  the Ricci curvature scalar $R$,
the gravitational Newton constant $G$, and the cosmological constant $\Lambda$ become dimensionless, $[G]=[R]=[\Lambda]=1$. 
Also the higher order gravitational terms, such as $R^2$ and $R^{\mu\nu}R_{\mu\nu}$, are dimensionless.

%%%%%%%%%%%%%%%%%%%%%%%%%%%%%%%%%%%%%%%%
%%%%%%%%%%%%%%%%%%%%%%%%%%%%%%%%%%%%%%%%
\begin{figure}[top]
\centerline{\includegraphics[width=0.7\linewidth]{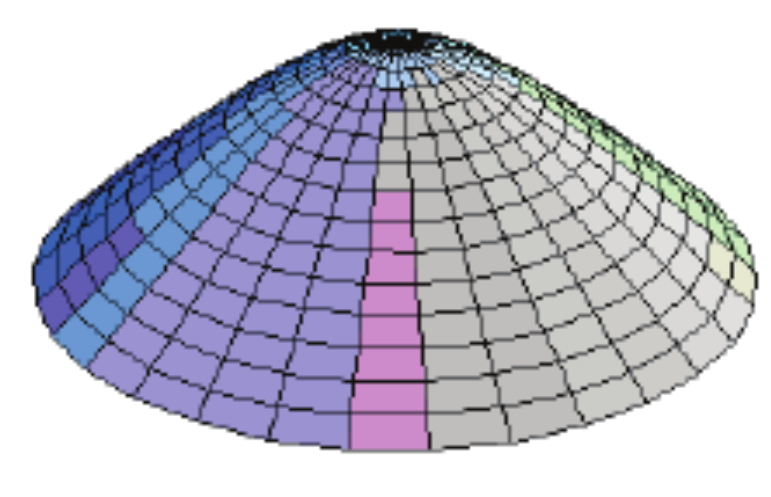}}
\label{HyperbolicFig} 
  \caption{2D illustration of the $3+1$ superplastic vacuum, in which the equilibrium size of the elementary cell is absent.
}
\end{figure}
%%%%%%%%%%%%%%%%%%%%%%%%%%%%%%%%%%%%%%%%
%%%%%%%%%%%%%%%%%%%%%%%%%%%%%%%%%%%%%%%%

Moreover, all the physical quantities become dimensionless. The reason for that is that  in the superplastic vacuum, which can be arbitrarily deformed, the equilibrium size of the elementary cell is absent, and thus the microscopic length scale (such as Planck scale)  is absent, see Fig. 11.
%\ref{HyperbolicFig}.
 That is why all the distances are measured in terms of the integer positions of the nodes in the crystal.

The hydrodynamics of systems with topological defects can be useful for the description of different types of spin and orbital glasses, observed in the superfluid phases of $^3$He in aerogel \cite{Eltsov2018}, see also Sec.\ref{LarkinDisorder}.

\section{Abrikosov-Beneslavskii-Herring monopole}
\label{ABHmonopoleSec}

The Polyakov hedgehog-monopole-instanton saga had one more important development, now extended to momentum space. The momentum-space counterpart of the Berry phase magnetic monopole is shown in Fig. 3% \ref{monopoles}
 ({\it right}). This Figure shows the topological signature of the $2\times 2$  Hamiltonian describing the Weyl fermions \cite{Volovik1987b}. The spin (or pseudospin in Weyl materials) forms the hedgehog in momentum space, representing the Berry phase monopole in momentum space. The stability of this hedgehog is also supported by topology, but now it is the topology in momentum space. The topological description of the band contact points can be found in the paper by Novikov \cite{Novikov1981}. Topological stability of the Wel point provides the emergence of the relativistic Weyl fermions in the vicinity of the hedgehog even in condensed matter nonrelativistic vacuum such as in superfluid $^3$He-A. Together with the chiral Weyl fermions, also gravity in terms of the tetrad fields and relativistic quantum gauge fields emerge in this superfluid. In other words, the whole Universe (or actually its caricature) can be found in a droplet of $^3$He \cite{Volovik2003}.

Unforunately at that time I was unaware on another, the essentially older, Universe created by Abrikosov, who together with Beneslavskii considered the relativistic Weyl fermions in semimetals \cite{Abrikosov1971,Abrikosov1972,Abrikosov1998}.
But in 1998, after the Abrikosov-70 workshop in Argonne, I got from Abrikosov the reference to his papers. Then I realized that as a student I visited his seminar talk in Chernogolovka in 1970, where  
he discussed the "relativistic" conical spectrum of electrons in semimetals. Though later I forgot about that seminar,  it was somehow deep in my subconscious mind. The Berry phase hedgehog-monopole in momentum space could be called Abrikosov-Beneslavskii-Herring (ABH) monopole. 

The topological invariant which describes the Weyl node can be also written in terms of the Green's function with the imaginary frequency \cite{GrinevichVolovik1988}. The Green's function has point singularity in the 4D momentum space $(\omega,p_x,p_y,p_z)$, see Eq.(\ref{TopInvariantMatrix}) in Sec. \ref{Yakovenko}.
It is the momentum-space analog of instanton. The description in terms of the Green's function is important in the case of strong interaction, when the single-particle Hamiltonian is ill defined.

\section{Gribov: RQFT in Majorana-Weyl superfluid $^3$He-A, Moscow zero, quark confinement}
\label{GribovSec}

During the years, Gribov's help was extremely important to me. Although he was usually very critical during seminars at the Landau Institute, he patiently answered my questions, perhaps because I did not belong to the high-energy community, and sometimes he even shared his ideas with me.
Gribov clarified to me different issues related to emergent RQFT in Weyl superfluid $^3$He-A. 
In one case I was puzzled by the logarithmically divering term in the action for $^3$He, which in terms of the effective $U(1)$ gauge field has the form $(B^2 - E^2) \ln [E_{\rm uv}^4/(B^2-E^2)]$, where $E_{\rm uv}$ is the ulraviolet cut-off (see Sec. \ref{KopninIordanskii} for definition of the synthetic gauge field below Eq.(\ref{Anomaly})). This term becomes imaginary for $E^2>B^2$, what to do with that? 
From the discussion with Gribov it became clear that this is nothing but the vacuum instability towards the Schwinger pair production, which occurs when the synthetic electric field exceeds the synthetic magnetic field.

The logarithmic divergence is the condensed matter analog of the zero charge effect -- the famous "Moscow zero"  by Abrikosov, Khalatnikov and Landau \cite{LandauAbrikosovKhalatnikov1954}. 
The zero charge effect is natural for the $U(1)$ gauge field.
However, to my surprize  the synthetic $SU(2)$ gauge field, which emerges in  $^3$He-A too, also obeys the zero charge behavior instead of the expected asymptotic freedom found by  Gross, Wilczek and Politzer
 \cite{GrossWilczekPolitzer1973}. The discussion with Gribov clarified this issue too. He simply asked me the question: "How many bosons and fermions do you have in your helium?"  It appeared that the number of the fermionic species is small compared to the number of bosonic fields. However, the fermions dominate because of the much larger ultraviolet cut-off $E_{\rm uv}$.

Gribov explained  to me also the origin of the Wess-Zumino term in the hydrodynamic action, and some other things. All this resulted in the paper on anomalies in $^3$He-A \cite{Volovik1987a} with acknowledgement to Gribov for numerous and helpful
discussions. The term in Eq.(4.9) there appeared after Gribov shared with me his view on the problem of zero charge effect for massless fermions, and then I realized that the same term with the same prefactor exists in superfluid hydrodynamics of $^3$He-A. It is the  Eq.(25) in the Gribov paper \cite{Gribov1987}.

Gribov was also the first one to  tell me that Bogoliubov quasiparticles in $^3$He-A are Majorana fermions. The zero energy modes found by Kopnin in the core of the $^3$He-A vortex \cite{KopninSalomaa1991}, see also \cite{MisirpashaevVolovik1995} and Sec. \ref{KopninMajorana},  appeared to be Majoranas.  Later I checked that the zero energy level does not shift from zero even in the presence of impurities 
\cite{Volovik1999a}, which is an important signature of the Majorana nature of the mode.

I continued the discussions with Gribov, even after he moved to Hungary, in particular in relation to the quark cofinement. My proposal to explain the confinement in terms of the ferromagnetic quantum vacuum, after discussion with Gribov  appeared to be the modification of the idea of the Savvidy vacuum \cite{Savvidy1977}.

More recently, with Klinkhamer, we tried to  extend the  Gribov picture  of
confinement  in QCD as diverging mass at low $k$ \cite{Gribov1978}. We came to the following estimation for the vacuum energy density (cosmological constant $\Lambda$)  in the present epoch \cite{KlinkhamerVolovik2009c}:
\begin{equation}
\Lambda=\rho_{\rm vac} \sim H \Lambda^3_{\rm QCD} \,,
\label{Gribov}
\end{equation}
where $H$ is the Hubble parameter, and $\Lambda_{\rm QCD}$ is the QCD energy scale.
The linear dependence on $H$ obtained in the phenomenological theory of confinement, has been also proposed in other approaches to QCD \cite{UrbanZhitnitsky2010,Zhitnitsky2014}.
 For the de-Sitter universe one has $\Lambda \sim H^2 E_{\rm Planck}^2$, which gives
\begin{equation}
\Lambda \sim \frac{\Lambda^6_{\rm QCD}}{E_{\rm Planck}^2}  \,.
\label{Zeldovich}
\end{equation}
Unfortunately the helpful  Gribov criticism is now missing.

On the other hand, since $\Lambda_{\rm QCD}$ is on the order of proton mass $m_p$, the equation (\ref{Zeldovich})
corresponds to the early suggestion by  Zeldovich, $\Lambda \sim Gm_p^6$  \cite{Zeldovich1967} (see also recent paper by Kamenshchik, Starobinsky and co-authors on the Pauli-Zeldovich mechanism of cancellation of the vacuum energy divergences \cite{KamenshchikStarobinsky2018}).

\section{Gor'kov: unconventional superconductivity, Weyl points and Dirac lines}
\label{Gorkov}

%%%%%%%%%%%%%%%%%%%%%%%%%%%%%%%%%%%%%%%%
%%%%%%%%%%%%%%%%%%%%%%%%%%%%%%%%%%%%%%%%
\begin{figure}[top]
\centerline{\includegraphics[width=0.6\linewidth]{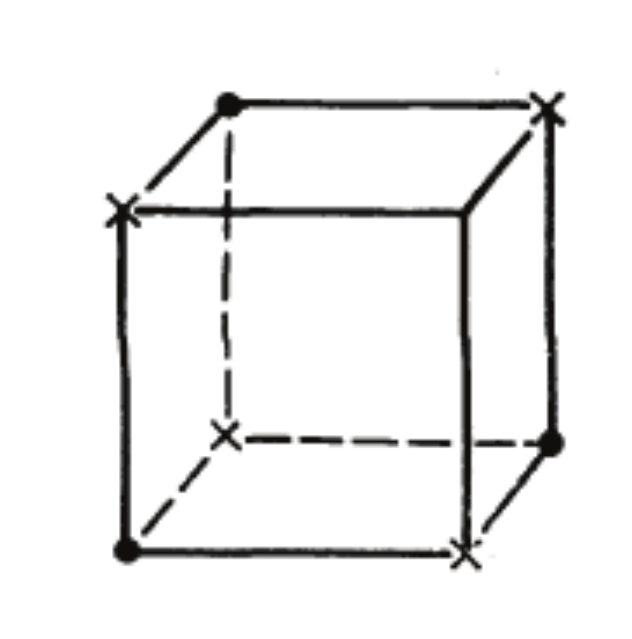}}
\label{Cube} 
  \caption{Figure from Ref. \cite{VolovikGorkov1985}. Arrangement of the nodes in the energy spectrum in superconductors of class $O(D_2)$. The points denote
four Weyl nodes  with topological charge $N=+1$ (the winding number of the hedgehog with spins outwards), and crosses denote four Weyl nodes  with $N=-1$  (the winding number of the hedgehog with spins inwards). 
In the vicinity of each Weyl node with $N=+1$ the chiral right-handed Weyl fermions emerge, while $N=-1$ is the topological charge of the left-handed quasiparticles. This arrangements of Weyl nodes can be compared with 8 right-handed and 8 left-handed particles (quarks and leptons) in each generation of Standard Model fermions.
 }
\end{figure}
%%%%%%%%%%%%%%%%%%%%%%%%%%%%%%%%%%%%%%%%
%%%%%%%%%%%%%%%%%%%%%%%%%%%%%%%%%%%%%%%%

Our collaboration with Gor'kov started when he returned from a conference, where new heavy-fermion superconductors had been discussed. Our collaboration led to the symmetry classification of the superconducting states \cite{VolovikGorkov1984,VolovikGorkov1985}.
Most of the unconventional superconducting states have nodes in the energy spectrum: Weyl points,  Dirac points and Dirac nodal lines.
One of the configurations with 8 Weyl points (4 right and 4 left)  is in Fig. 12.
%\ref{Cube}. 
The extension of this configuation to 4D space produces the 4D analog of graphene \cite{Creutz2008,Creutz2014}, with 8 left and 8 right Weyl fermions, as in each generation of the Standard Model fermions  (see also the paper in memory of Gor'kov \cite{VolovikOnGorkov2017}, where the superconducting state with 48 Weyl fermions is discussed).

%%%%%%%%%%%%%%%%%%%%%%%%%%%%%%%%%%%%%%%%
%%%%%%%%%%%%%%%%%%%%%%%%%%%%%%%%%%%%%%%%
\begin{figure}[top]
\centerline{\includegraphics[width=\linewidth]{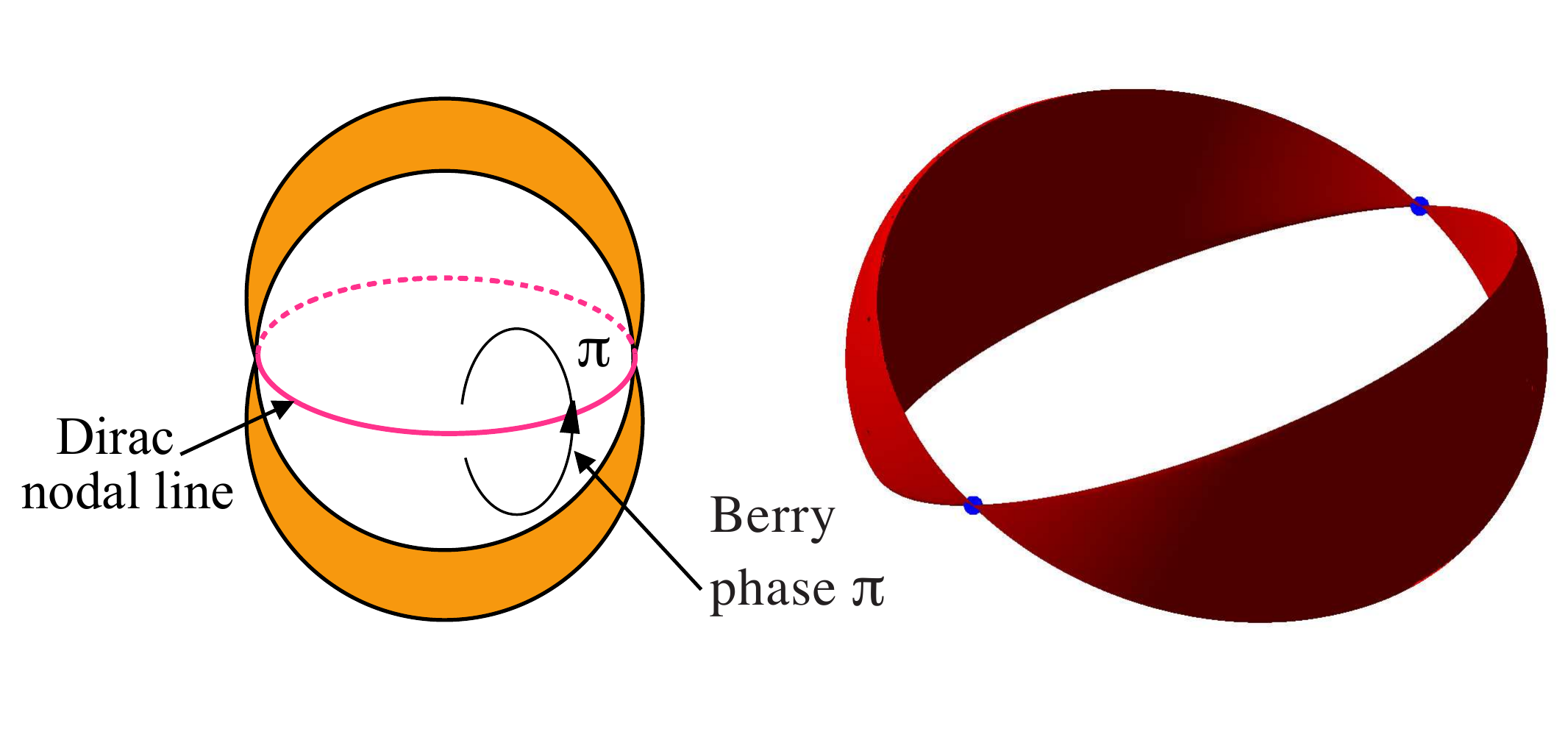}}
\label{Superflow} 
  \caption{Nodal line in the polar phase of $^3$He ({\it left}) and its transformation to Bogoliubov Fermi surface under superflow  ({\it right}).
}
\end{figure}
%%%%%%%%%%%%%%%%%%%%%%%%%%%%%%%%%%%%%%%%
%%%%%%%%%%%%%%%%%%%%%%%%%%%%%%%%%%%%%%%%

The Dirac lines also appear in many classes of superconductivity, including cuprate superconductors. The Dirac line has an important effect on the thermodynamics of superconductors \cite{Volovik1993b}. The reason is that in the presence of a supercurrent, the nodes in the spectrum transform to 
Fermi surfaces, see Fig. 13.
% \ref{Superflow}.
%\ref{Superflow}. 
Such Fermi surfaces emerge in superconductors due to broken time reversal symmetry or parity, and now they are called the Bogoliubov Fermi surfaces \cite{Agterberg2018}. The Bogoliubov Fermi surface provides the nonzero density of states (DoS), which in the case of the nodal line  is proportional to the superfluid velocity. For the Abrikosov  vortex lattice the flow around vortices produces a DoS  which is proportional to $\sqrt{B}$ \cite{Volovik1988d,Volovik1993b}. Gor'kov called this phenomenon "koreshok"  -- the diminutive form of the Russian word  koren' (root).

\section{Dzyaloshinskii, Polyakov and Wiegmann paper and $\theta$-term in $^3$He-A}

The paper by Dzyaloshinskii, Polyakov and Wiegmann \cite{DzyaloshinskiiPolyakovWiegmann1988} inspired the work on the possible $\theta$-term in thin films of chiral superfluid $^3$He-A \cite{Volovik1988c}. The consequences of that are the intrinsic quantum Hall effect, spin quantum Hall effect,
and exotic spin and statistics of solitons, which depend on film thickness
\cite{VolovikSolovyevYakovenko1989}. These works were made under extremely useful  discussions with Wiegmann.

\section{Yakovenko, Grinevich and topology in momentum space}
\label{Yakovenko}

During our collaboration with Yakovenko, we expressed the intrinsic quantum Hall and spin quantum Hall effects via the $\pi_3$ topological Chern numbers in terms of the Green's function \cite{VolovikYakovenko1989}.
The same invariants, but where the integral is around the Weyl point in the 4D $p_\mu$ space:
\begin{equation}
N =
\frac{1}{{24\pi^2}}e_{\mu\nu\lambda\gamma}{\bf tr} ~
 ~\int_{\sigma}~  dS^{\gamma}
~ {\cal G}\partial_{p_\mu} {\cal G}^{-1}
{\cal G}\partial_{p_\nu} {\cal G}^{-1} {\cal G}\partial_{p_\lambda}  {\cal
G}^{-1} \,,
\label{TopInvariantMatrix}
\end{equation}
have been used in our paper with Grinevich for description of the topological protection of the Weyl fermions \cite{GrinevichVolovik1988}.
Here $\sigma$ is the closed 3D surface around the point in 4D momentum-frequency space.
The value of this Chern number in Eq.(\ref{TopInvariantMatrix}) is equal to the charge of the ABH monopole in Fig. 2
%\ref{monopoles}
 ({\it right}).
It is the instanton description of the Polyakov hedgehog-monopole in momentum space, 
see Sec. \ref{ABHmonopoleSec}.

\section{Larkin and disorder}
\label{LarkinDisorder}

%%%%%%%%%%%%%%%%%%%%%%%%%%%%%%%%%%%%%%%%
%%%%%%%%%%%%%%%%%%%%%%%%%%%%%%%%%%%%%%%%
\begin{figure}[top]
\centerline{\includegraphics[width=\linewidth]{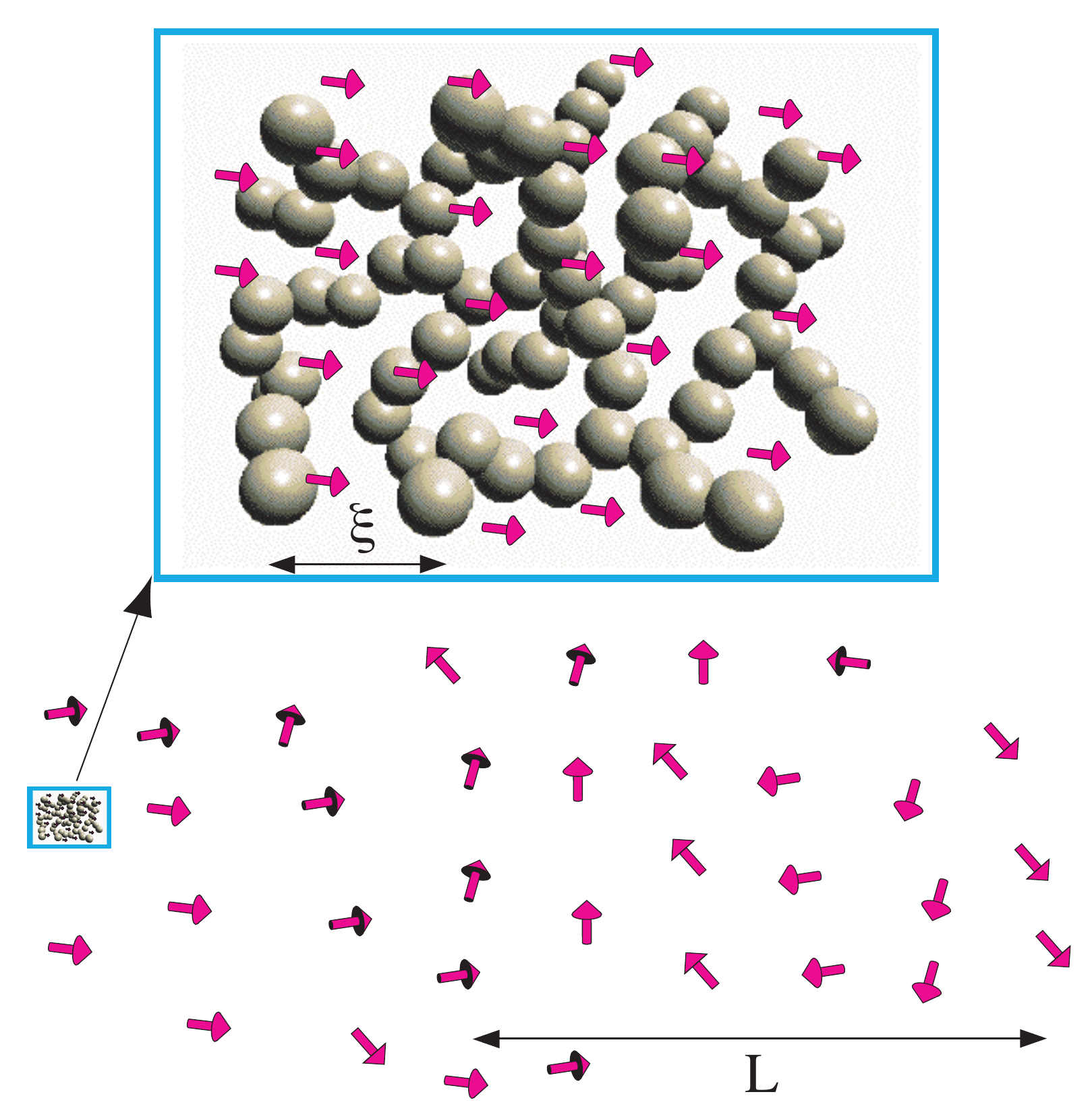}}
\label{LIMstate} 
  \caption{Larkin-Imry-Ma state of $^3$He-A in aerogel. Long-range orientational order is destroyed by  weak interaction with aerogel strands, which provide the random anisotropy. Here $L$ is the legth scale at which the long-range order is destroyed. It is much bigger than the characteristic scale of the quenched random anisotropy.
}
\end{figure}
%%%%%%%%%%%%%%%%%%%%%%%%%%%%%%%%%%%%%%%%
%%%%%%%%%%%%%%%%%%%%%%%%%%%%%%%%%%%%%%%%

A suprising result of Larkin is that even small disorder destroys the Abrikosov vortex lattice
\cite{Larkin1970}. In magnets the similar effect is the destruction of orientational long-range order by  weak random anisotropy \cite{ImryMa}. 
Following this trend in the Landau Institute, it was suggested that a similar effect can be realized in $^3$He-A in aerogel, where the weak random anisotropy provided by the disordered aerogel strands may destroy the  long-range orientational order \cite{Volovik1996}, see Fig. 14.
%\ref{LIMstate}.
 This disordered state, which we called the Larkin-Imry-Ma state,  has been observed in NMR experiments \cite{Dmitriev2010,Askhadullin2015}, which opened the route to study experimentally many different types of spin and orbital glasses in superfluid $^3$He
\cite{Eltsov2018}.

\section{Kopnin and Iordanskii forces}
\label{KopninIordanskii}

%%%%%%%%%%%%%%%%%%%%%%%%%%%%%%%%%%%%%%%%
%%%%%%%%%%%%%%%%%%%%%%%%%%%%%%%%%%%%%%%%
\begin{figure}[top]
\centerline{\includegraphics[width=\linewidth]{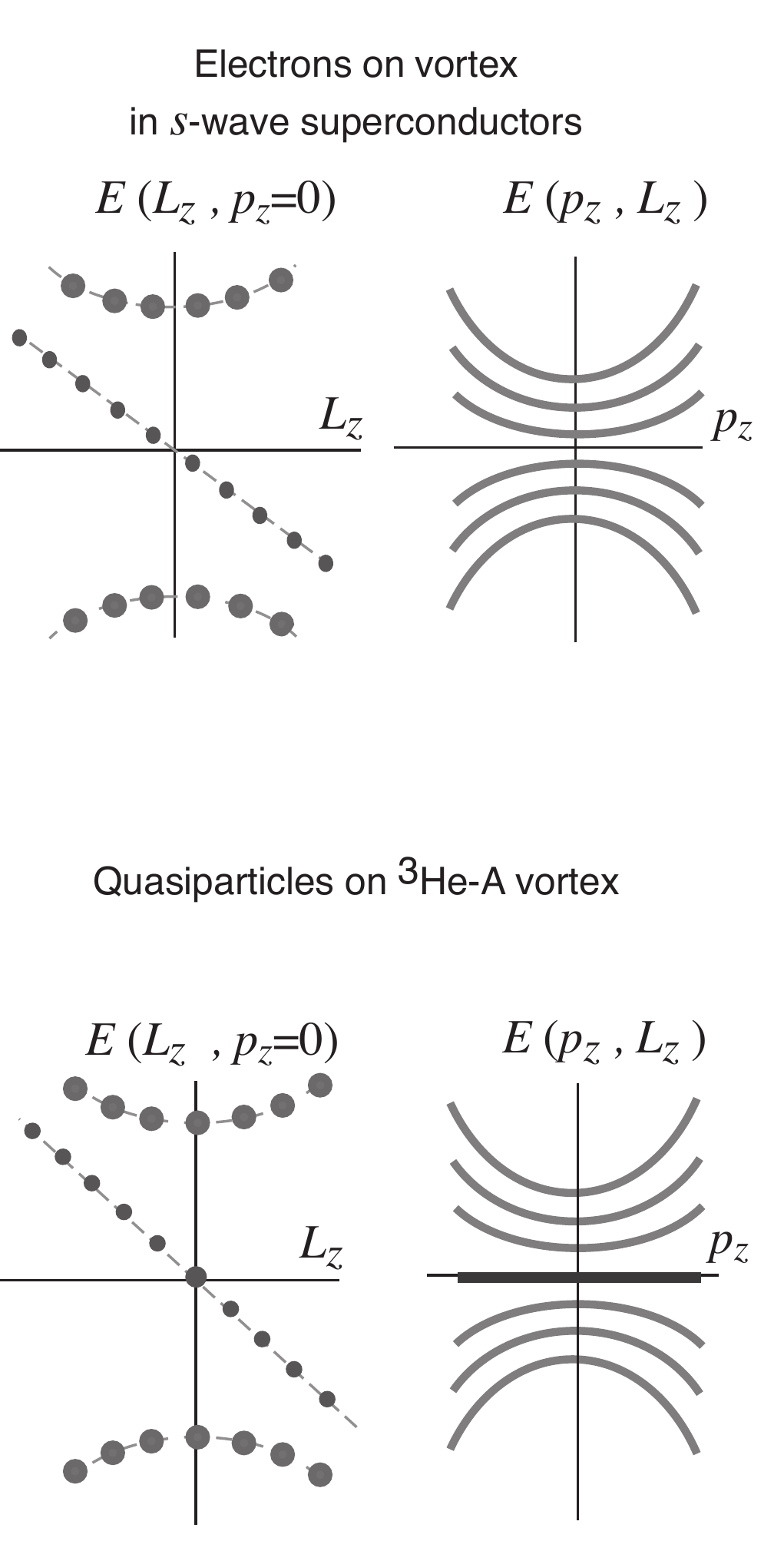}}
\label{FermionOnVortex} 
  \caption{
({\it top left}): Chiral branch of fermions living in the core of Abrikosov vortices in $s$-wave superconductors. The spectrum is the function of discrete angular momentum $L_z$. The spectral flow along this anomalous branch is the origin of the extra force acting on a vortex -- the Kopnin force. ({\it top right}): The spectrum as function of $p_z$ at fixed values of $L_z$. The spectrum has a minigap of the order $\omega_0=\Delta^2/E_F$.
({\it bottom left}):  Chiral branch of fermions living in the core of  vortices in chiral $p$-wave superfluid. This spectrum contains Majorana zero energy mode. ({\it bottom right}): The spectrum as function of $p_z$ in superfluid $^3$He-A, where the branch with $L_z=0$ represents the flat band of Majorana fermions.
 }
\end{figure}
%%%%%%%%%%%%%%%%%%%%%%%%%%%%%%%%%%%%%%%%
%%%%%%%%%%%%%%%%%%%%%%%%%%%%%%%%%%%%%%%%

Starting with 1981,  I had collaboration with the experimental team in Helsinki, which studied different types of vortices and other topological defects in a unique rotating cryostat operating at milliKelvin temperatures. For this I had to study the vortex dynamics which at that time has been developed by Kopnin in superconductors.

At first glance the Kopnin theory of vortex dynamics \cite{Kopnin1976,Kopnin2001} looked rather complicated. Fortunately, it happened that his theory could be reformulated in more simple terms. The vortex represents the chiral system, and the spectrum of the fermionic modes localized in the vortex core has the anomalous branch which as function of the discrete angular momentum $L_z$ "crosses" zero, see 
Fig. 15
%\ref{FermionOnVortex} 
({\it top left}) and  ({\it  bottom left}). It appears that the spectral flow along this anomalous branch is responsible for the Kopnin force acting on the vortex. Thus the Kopnin force can be explained in terms of the chiral anomaly as the analog of the Callan-Harvey effect \cite{Volovik1993c}. The chirality here is generated by the vorticity: the number of the anomalous branches, which "cross" zero, is determined by the vortex winding number $N$. 

The Kopnin spectral flow force adds to the conventional Magnus force acting on the vortex, which exists in conventional liquids, and to Iordanskii force already well known in two fluid dynamics of superfluids \cite{Iordanskii1964,Iordanskii1966}. The Kopnin spectral flow force is of fermionic origin and exists only in fermionic superfluids and in superconductors.

%%%%%%%%%%%%%%%%%%%%%%%%%%%%%%%%%%%%%%%%
%%%%%%%%%%%%%%%%%%%%%%%%%%%%%%%%%%%%%%%%
\begin{figure}[top]
\centerline{\includegraphics[width=\linewidth]{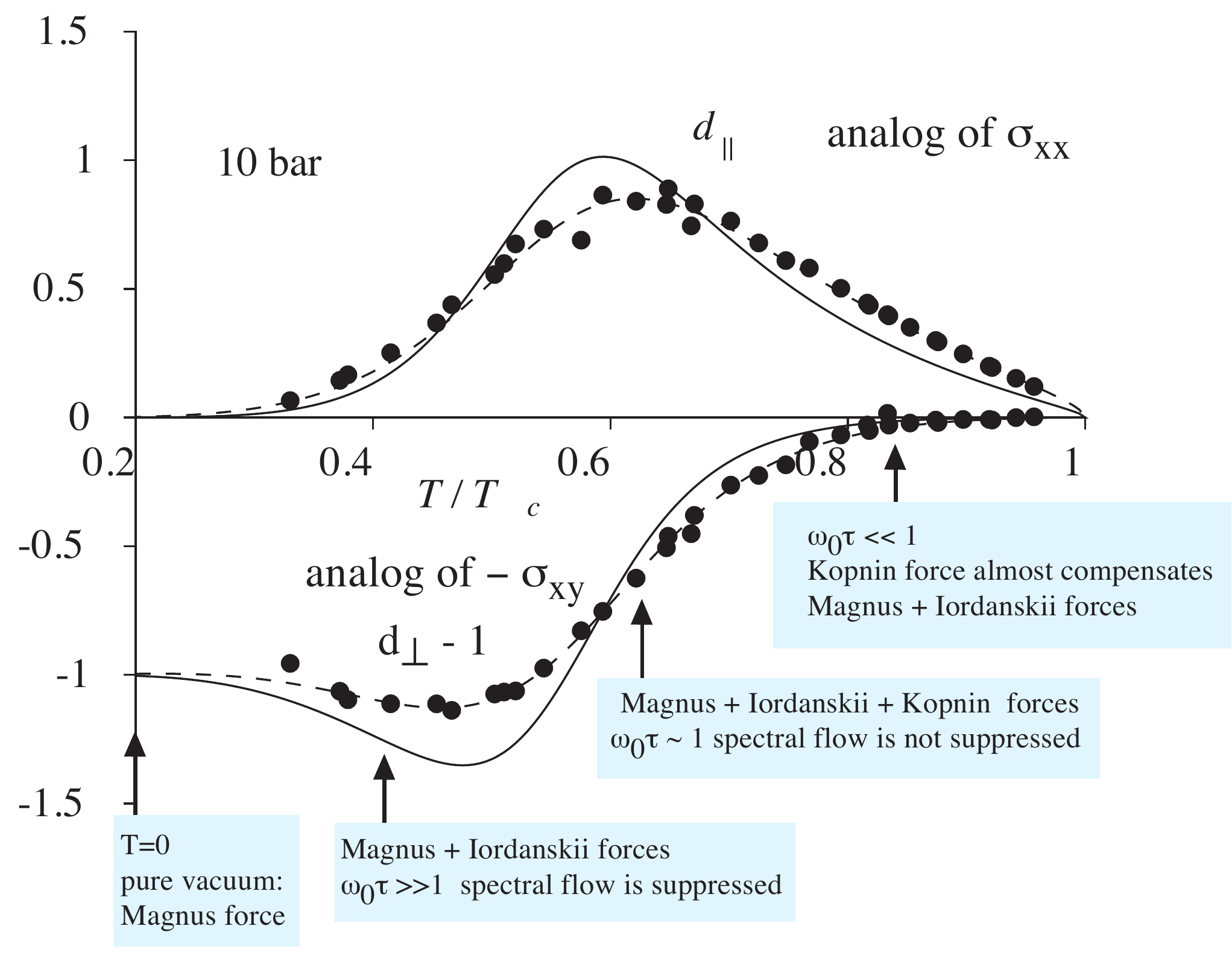}}
\label{3forces} 
  \caption{Measurement of three nondissipative forces  acting on quantized vortices in rotating $^3$He-B: the Magnus force known in conventional liquids, Iordanskii force known in two-fluid dynamics of superfluids, and Kopnin spectral flow force, which exists only in Fermi superfluids and comes from the analog of chiral anomaly for  fermions
living in the vortex core. The solid line is Kopnin calculations.
 }
\end{figure}
%%%%%%%%%%%%%%%%%%%%%%%%%%%%%%%%%%%%%%%%
%%%%%%%%%%%%%%%%%%%%%%%%%%%%%%%%%%%%%%%%

After the origin of Kopnin force was clarified, our cooperation with Kopnin on 
vortex dynamics was developed \cite{Kopnin1995,Kopnin1997}.
Finally, the Kopnin theory has been confirmed in experiments on vortices in $^3$He-B \cite{Bevan1997}: the measured temperature dependence of the Kopnin force agreed with his calculations. Note that superfluid $^3$He does not contain impurities, and vortices are not pinned. This allowed us to measure the Kopnin, Iordanskii and Magnus forces in their pure form, see Fig. 16.
%\ref{3forces}.

In case of continuous vortices in $^3$He-A -- skyrmions --  the Kopnin force can be fully described by the Adler-Bell-Jackiw equation for chiral anomaly:
\begin{equation}
\partial_\mu J^\mu_5=\frac{1}{4\pi^2} q^2{\bf B}\cdot {\bf E} \,,
\label{Anomaly}
\end{equation}
which confirms the chiral anomaly origin of the Kopnin force in the general case.
In Eq.(\ref{Anomaly}), the synthetic gauge fields comes from the time and space dependence of the position of the node ${\bf K}({\bf r},t)$ in the spectrum of Weyl quasiparticles in the presence of the moving skyrmions: ${\bf B}=\nabla\times {\bf K}$ and  ${\bf E}=\partial_t {\bf K}$.
In the weak coupling approximation, the Kopnin force compensates the Magnus force practically at any temperature, which  has been confirmed in experiments on vortices in $^3$He-A \cite{Bevan1997}. 

The Kopnin force was also very important in study of turbulence in the flow of superfluid $^3$He.
Since the Kopnin force has similar dependence on velocity as the mutual friction force between the normal component of the liquid and quantized vortices, the corresponding Reynolds number ${\cal R}(T)=\omega_0\tau$ does not depend on velocity and is only determined by temperature, where $\omega_0$ is the minigap and $1/\tau$ is the width of the vortex core levels.
 The transition from the laminar to turbulent flow  takes place at temperature when ${\cal R}(T)\sim 1$. Such a transition
governed by this novel Reynolds number has been experimentally observed, see the review \cite{Finne2006}.

\section{Kopnin, Majorana fermions  and flat band superconductivity}
\label{KopninMajorana}

A very interesting result obtained by Kopnin concerns the fermion modes living in the core of the singular $N=1$ vortex in  chiral superfluid $^3$He-A in Fig. 15
%\ref{FermionOnVortex}  
({\it bottom}) \cite{KopninSalomaa1991,MisirpashaevVolovik1995,Volovik1999a}. It was found that the branch of the spectrum with zero angular momentum $L_z=0$ is dispersionless, $E_0(p_z)=0$ in some region of momenta, $-p_F<p_z<p_F$,
see  Fig. 15
%\ref{FermionOnVortex}
 ({\it bottom}). 
This observation inspired to look for the topological origin of this 1D flat band with zero energy. It appeared that in the 2D case the state with exactly zero energy  corresponds to the Majorana mode on the vortex \cite{Volovik1999a,Ivanov2001}.
In the 3D case, the existence of the 1D Majorana flat band is connected to the existence of the Weyl nodes in bulk: the boundaries of the flat band are determined by the projections of the Weyl nodes in bulk to the vortex line  \cite{Volovik2011c}.
This is one of the many examples of bulk-boundary and bulk-vortex correspondence in topological materials.

%%%%%%%%%%%%%%%%%%%%%%%%%%%%%%%%%%%%%%%%
%%%%%%%%%%%%%%%%%%%%%%%%%%%%%%%%%%%%%%%%
\begin{figure}[top]
\centerline{\includegraphics[width=0.8\linewidth]{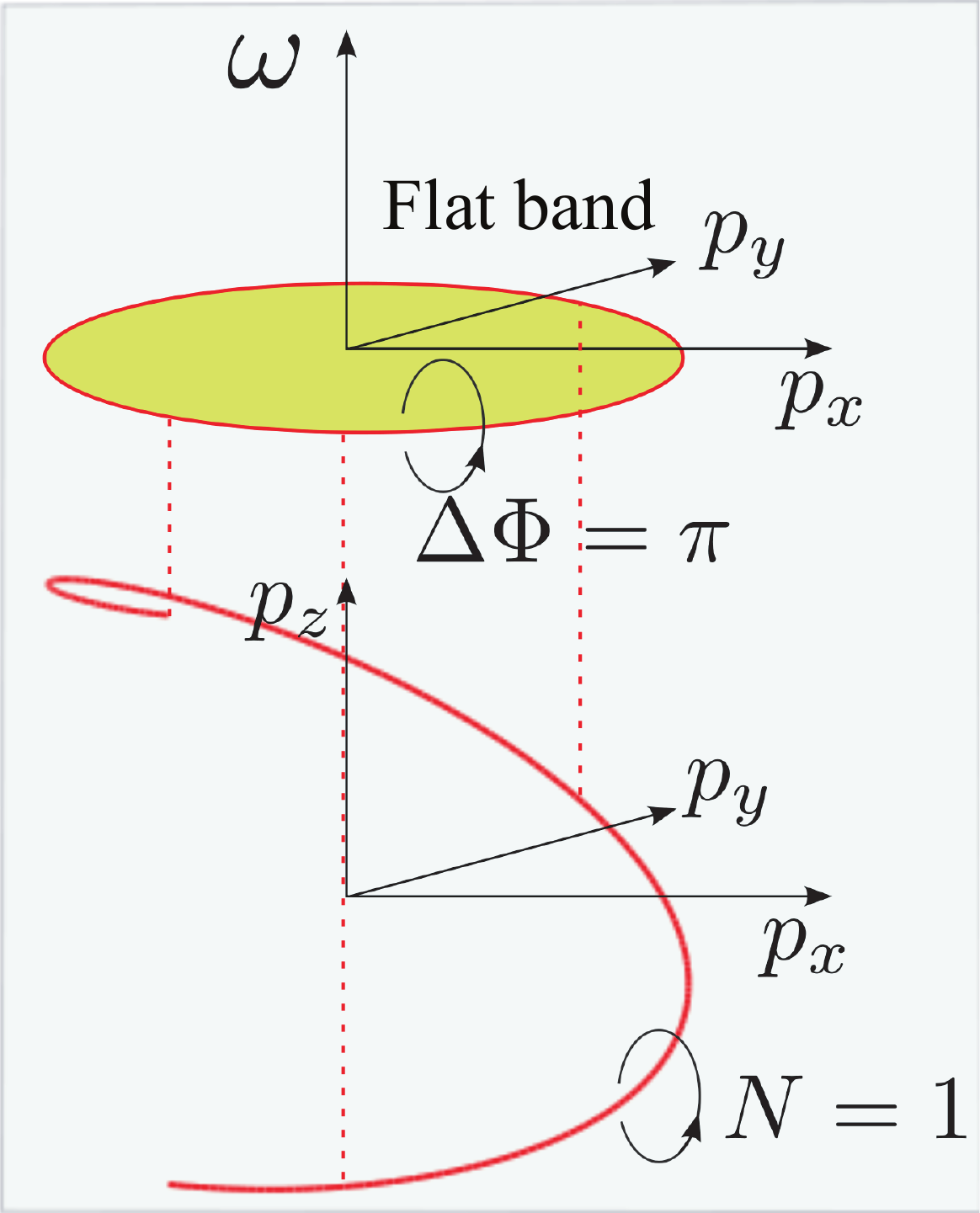}}
\label{FlatBand} 
  \caption{Bulk-surface correspondence in semimetals with nodal line in bulk. Dirac nodal line  gives rise to flat band on the surface. The boundary of the flat band coincides with the projection of the Dirac line to the surface. 
 }
\end{figure}
%%%%%%%%%%%%%%%%%%%%%%%%%%%%%%%%%%%%%%%%
%%%%%%%%%%%%%%%%%%%%%%%%%%%%%%%%%%%%%%%%

Even more important example is the 2D topological flat band on the surface of semimetals having nodal lines in the bulk spectrum. The boundary of the 
surface flat band is determined by  projection of the nodal line to  the surface of the semimetal \cite{HeikkilaVolovik2011}, see Fig. 17.
%\ref{FlatBand}. 
The singular density of states in the flat band leads to
the flat band superconductivity
\cite{Kopnin2011,Kopnin2011b,Kopnin2013b}. The flat band superconductivity is characterized by the linear dependence of the transition temperature on the interaction $g$ in the Cooper 
pair channel, $T_c \sim g V_{\rm FB}$, where  $V_{\rm FB}$ is the volume or the area of the flat band \cite{Khodel1990}. This is in contrast to the conventional superconductvity in metals with Fermi surfaces, where $T_c$ is exponentially suppressed.

Recently superconductivity has been observed  in the twisted bilayer 
graphene\cite{Cao2018a,Cao2018b}.  The maximum of $T_c$ takes place at the "magic angle" of twist, at which the electronic band structure becomes nearly flat, see discussion in Ref. \cite{Volovik2018a} and 
 Ref.\cite{Hekkila2018} (Fig. 18).
%\ref{TwistedGraphene}).

%%%%%%%%%%%%%%%%%%%%%%%%%%%%%%%%%%%%%%%%
%%%%%%%%%%%%%%%%%%%%%%%%%%%%%%%%%%%%%%%%
\begin{figure}[top]
\centerline{\includegraphics[width=\linewidth]{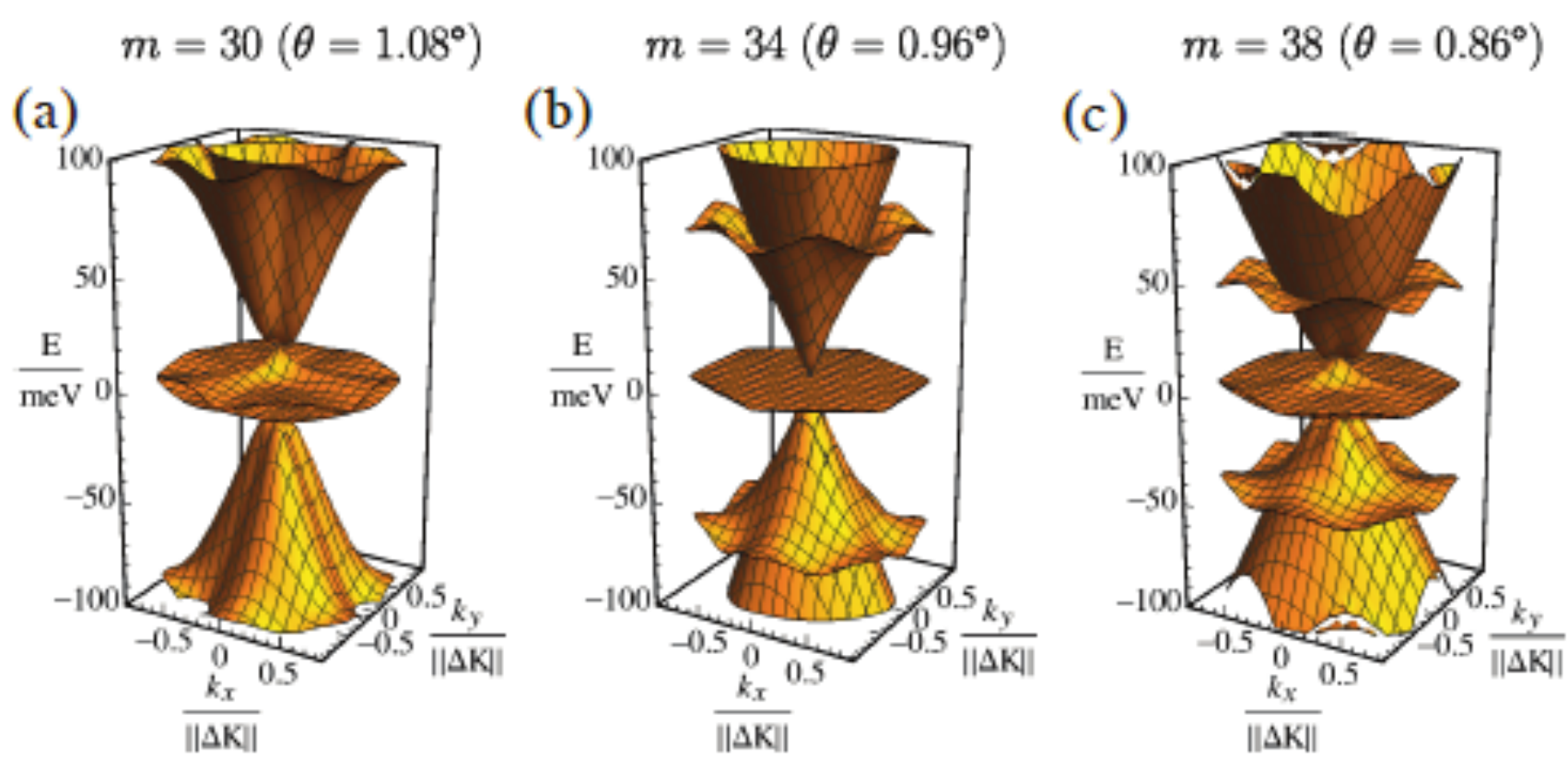}}
\label{TwistedGraphene} 
  \caption{ Spectrum of electrons in twisted bilayer graphene, where the flat band emerges at magic angle of twist (from Ref. \cite{Hekkila2018}).
}
\end{figure}
%%%%%%%%%%%%%%%%%%%%%%%%%%%%%%%%%%%%%%%%
%%%%%%%%%%%%%%%%%%%%%%%%%%%%%%%%%%%%%%%%

For vortices in superfluids and superconductors with nodal lines in bulk, the singularities in thermodynamics come from the regions far away from the vortex, see the discussion on "koreshok" in Sec. \ref{Gorkov}. For cuprate superconductors they are discussed in \cite{Volovik1993b} and in our paper with Kopnin \cite{Kopnin1996}.

\section{Fomin, coherent precession, magnon BEC}

%%%%%%%%%%%%%%%%%%%%%%%%%%%%%%%%%%%%%%%%
%%%%%%%%%%%%%%%%%%%%%%%%%%%%%%%%%%%%%%%%
\begin{figure}[top]
\centerline{\includegraphics[width=\linewidth]{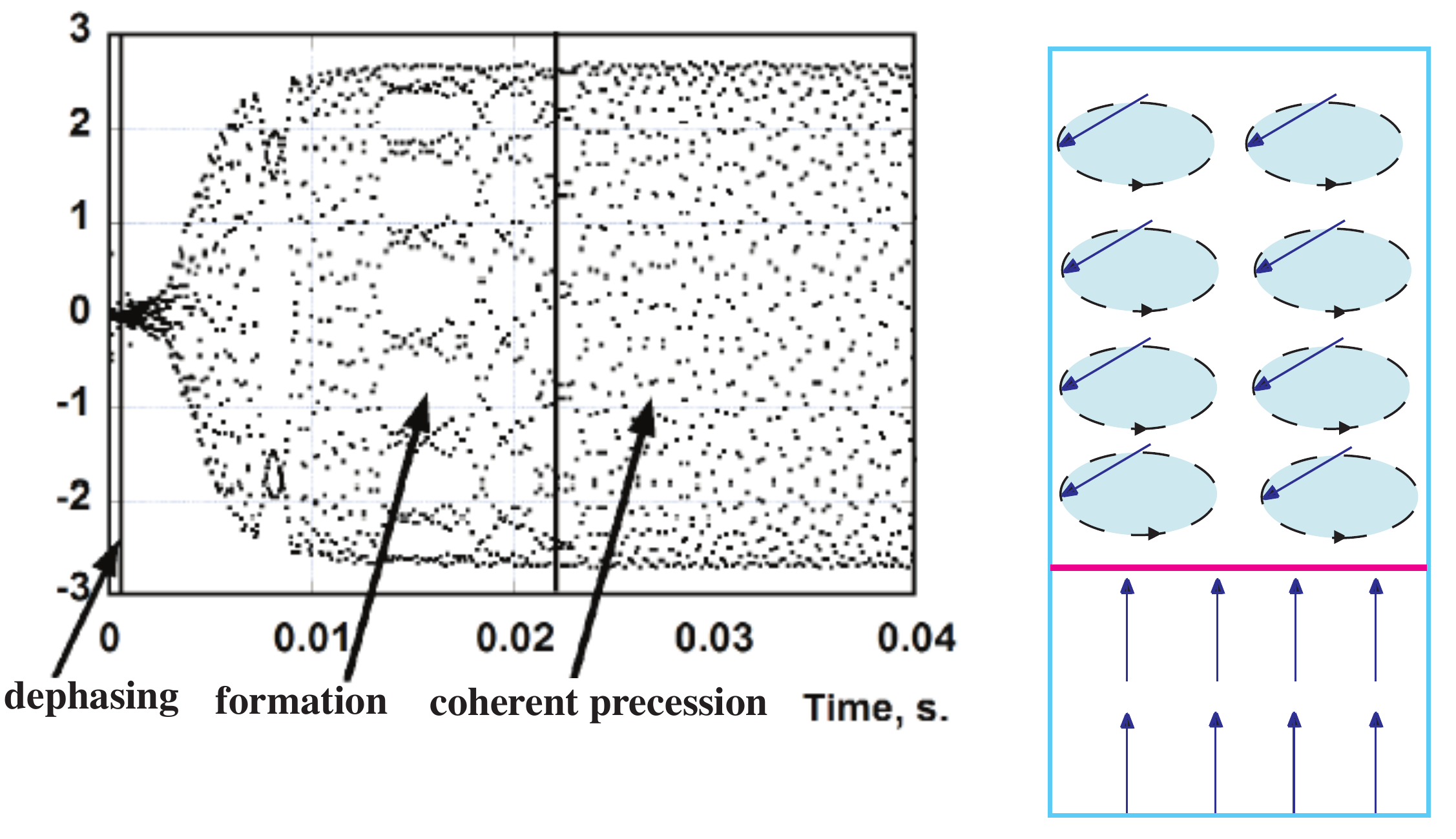}}
\label{MagnonBEC} 
  \caption{ Spontaneously formed coherent precession of magnetization in $^3$He-B discovered in Kapitza Institute in collaboration with Fomin. ({\it left}):  After strong pulse of radio-frequency field, which deflects magnetization on large angle, the spins start to precess around external magnetic field. Due to inhomogeneity of the system the spins precess with different frequences, which leads to dephasing, and the measured signal completely disappears. Then the miracle occurs: without any external drive the coherent dynamical state is spontaneously formed, in which all the spins precess with the same collective frequency and with the same phase ignoring inhomogeneity of the system. The precessing state is concentrated  in the part of the cell  ({\it right}), which was called Homogeneously Precessing Domain (HPD). 
The precession is slowly decaying due to nonconservation of spin, but during the decay it remains phase coherent, only the volume of the precessing domain slowly decreases with time.
}
\end{figure}
%%%%%%%%%%%%%%%%%%%%%%%%%%%%%%%%%%%%%%%%
%%%%%%%%%%%%%%%%%%%%%%%%%%%%%%%%%%%%%%%%

Fomin got the London Prize together with experimentalists from Kapitza Insitute, Bunkov and Dmitriev, for the discovery of the spontaneously formed coherent precession of magnetization in superfluid $^3$He-B \cite{Fomin1984,Fomin1985,HPDexp}, see Fig. 19.
%\ref{MagnonBEC}. 
This is a unique example  of spontaneous self-organization in a quantum system. Spins, which were originally precessing with different frequences, form   the collective state in which all the spins precess with the same  frequency and with the same phase. This state lives for a long time without external drive and inspite of the inhomogeneity of the system. 

For me, this phenomenon was not very clear. However, it worked. Due to connection between Kapitza Institute and Low Temperature Lab in Helsinki, this phenomenon was brought to Helsinki, where the HPD appeared to be very useful as a tool for experimental investigation of topological defects in $^3$He-B -- vortices and solitons. In particular, using HPD the exotic topological object -- 
combined spin-mass vortex with soliton tail -- has been observed and identified  \cite{Kondo2018}.
So, I had to study this phenomenon in detail.

Again, as in the case with Kopinin theory of vortex dynamics, the Fomin theory of HPD was beautiful, but it was very difficult for me to apply it to our problems. And again, it happened that Fomin's theory could be reformulated in a more simple way: in terms of the Bose-Einstein condensate (BEC) of quasiparticles -- magnons \cite{Volovik2008}, see also review \cite{BunkovVolovik2013}. 
The reason for that is that the coherent precession has the off-diagonal long-range order (ODLRO) signature similar to that in conventional superfluid:
\begin{equation}
S_x + iS_y = \left<\hat{\bf S}^+\right> = S\sin \beta e^{i(\alpha + \omega t)}
\,,
 \label{ODLROspin}
\end{equation}
where ${\bf S}^+$ is the operator of creation of spin and $\beta$ is the tipping angle of magnetization.
Using Holstein-Primakoff transformation one can rewrite this in terms of magnon BEC, where the order parameter is the quasi-average of the operator of annihilation of magnon number:
\begin{equation}
\Psi = \left<\hat{\mbox{\boldmath$\Psi$}}\right> = \sqrt{\frac{2S}{\hbar}} \sin \frac{\beta}{2}  e^{i(\alpha + \mu t)}
\,,
 \label{ODLROmagnon}
\end{equation}%
The role of the global phase of precession $\omega$ is played by the chemical potential $\mu$ of the pumped magnons, which is well defined if the slow spin relaxation is neglected. 

The close connection between the long-lived coherent precession of spins and the superfluid/superconducting states with the off-diagonal long-range order is also supported by observation that  superconductivity can be represented as the coherent precession of Anderson pseudospins.\cite{Anderson1958} However, as distinct from the long-lived quasi-equilibrium coherent precession of conventional spins, superconductivity is the true equilibrium phenomenon. The reason for that is that the proper projection of the total Anderson pseudospin coincides with the number of electrons, and thus is fully conserved, as distinct from the quasiconservation of the magnon number.

If the spin relaxation is neglected, the dynamics of the precessing system can be determined by the corresponding Landau-Khalatnikov hydrodynamics, applied now to magnon superfluid. Then all the phenomena related to the coherent precession, old and new, could be described in the same way: spin current Josephson effect; magnon BEC in magnetic and textural trap; Goldstone mode of precession -- phonon in magnon BEC; 
magnonic analog of MIT bag model of hadrons \cite{Autti2012};
magnonic analog of relativistic $Q$-ball \cite{Autti2018b}; etc. 

The coherent precession has also some signatures of the so-called time crystal \cite{TimeQuasicrystal2018}.
If the spin-orbit interaction is neglected, the magnon number is conserved, and the precessing state is the ground state of the system with fixed number of magnons. So, we have the oscillations in the ground state,
as suggested by Wilczek \cite{Wilczek2013}. But  without spin-orbit interaction  these oscillations are not observable.

Theory of magnon BEC in solid state materials (yttrium iron garnet films) has been considered by Pokrovsky, see review 
\cite{Pokrovsky2017}.

\section{Polyakov, Starobinsky, cosmological constant and vacuum decay}

The counterpart of the Polyakov hedgehog-monopole in momentum space -- the Weyl point -- naturally gives rise to the emergent  gravitational field acting on Weyl fermions. This again sparked my interest in  topics related to gravity, but now on a more serious ground than the analogy with superfluid $^4$He in Sec. \ref{KhalatLecture}. 
In this respect the consultations with Starobinsky became highly important and  extremely useful.

One of the directions was black hole radiation, which was started  by Zel'dovich 
\cite{Zeldovich1971} and Starobinsky \cite{Starobinskii1973} for rotating black holes, and continued by Hawking for  nonrotating black holes. In Landau Institute, this issue has been rather popular: I can mention 
Belinski \cite{Belinski1995} with whom I had many discussions and Byalko \cite{Byalko1979}.

 It appeared that the Zel'dovich-Starobinsky radiation by a rotating black hole can be simulated by a body rotating in a superfluid vacuum \cite{Calogeracos1999,Takeuchi2008}, while the Hawking radiation   using superflow or moving texture \cite{JacobsonVolovik1998,Volovik1999}. Also both the Hawking radiation and the Zel'dovich-Starobinsky radiation can be described in terms of the semiclassical tunneling. 
The same semiclassical approach can be applied to the radiation from the de Sitter cosmological horizon. But the latter already touches a different direction -- the problems related to the vacuum energy and cosmological constant. In this direction, the Starobinsky inflation 
\cite{StarobinskyYokoyama1994,Kofman1997} is the key issue.

%%%%%%%%%%%%%%%%%%%%%%%%%%%%%%%%%%%%%%%%
%%%%%%%%%%%%%%%%%%%%%%%%%%%%%%%%%%%%%%%%
\begin{figure}[top]
\centerline{\includegraphics[width=\linewidth]{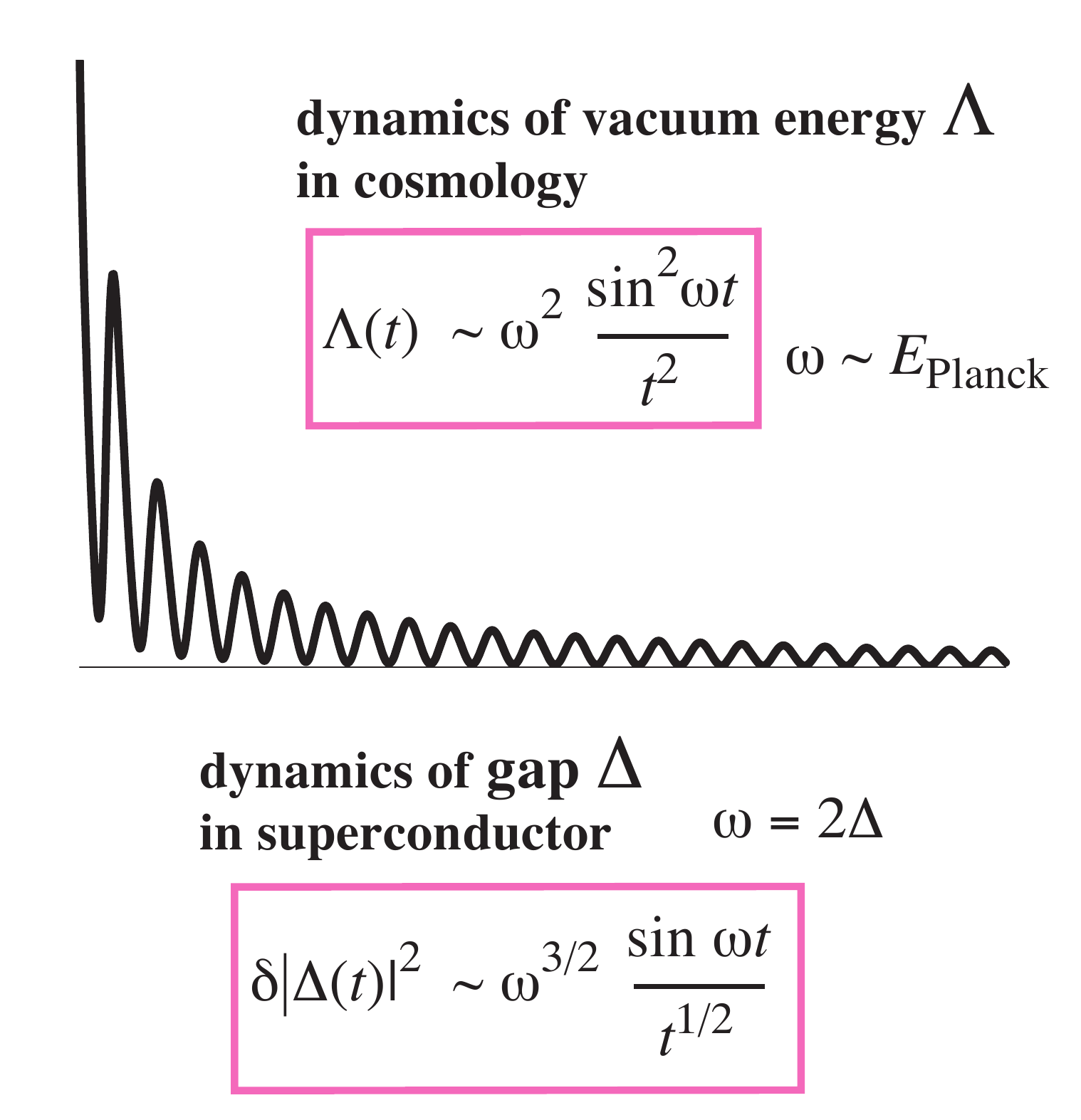}}
\label{VacDecay} 
  \caption{Processes of the vacuum decay after the Big Bang in the Universe and after the quench in superconductors and fermionic superfluids. In both cases the decay is accompanied by oscillations with frequency corresponding to the mass of the inflaton -- the Higgs field or the $q$-field. 
}
\end{figure}
%%%%%%%%%%%%%%%%%%%%%%%%%%%%%%%%%%%%%%%%
%%%%%%%%%%%%%%%%%%%%%%%%%%%%%%%%%%%%%%%%

In a series of papers with Klinkhamer \cite{KlinkhamerVolovik2008a,KlinkhamerVolovik2008b,KlinkhamerVolovik2008c}, we introduced the so-called $q$-theory, where the vacuum is described by a dynamical variable introduced by Hawking, 
the 4-form field \cite{Hawking1984+Duff+Wu}. The nonlinear extension of the Hawking theory allowed us to study the thermodynamics and dynamics of the quantum vacuum. 
The approach appeared to be rather general. Instead of the Hawking 4-form field one may use the other 
variables, which can describe the physical vacuum, but they lead to the same dynamical equations. One of such variable 
\cite{KlinkhamerVolovik2016} has been inspired by the papers by Kats and Lebedev  on a freely suspended film \cite{Kats2015}.
The main advantage of such an approach is that in full equilibrium the properly defined vacuum energy, which enters Einstein equations as cosmological constant,  is zero without fine tuning.  The mechanism of cancellation is purely thermodynamic and does not depend on whether the vacuum is relativisic or not. In this respect it is very different from the Pauli-Zeldovich mechanism discussed by Kamenshchik and Starobinsky \cite{KamenshchikStarobinsky2018}, which relies  on the cancellation of contributions of relativistic bosons and relativistic fermions.

 The thermodynamic approach solves the main cosmological constant problem: in  the Minkowski vacuum the huge vacuum energy is naturally cancelled. The problem remains, however, in the dynamics.
 If one assumes that the Big Bang has started in the originally equilibrium vacuum, then from our equations without dissipation it follows that 
the cosmological constant, which is very large immediately after the Big Bang, relaxes with oscillations and
its magnitude averaged over fast oscillations reaches the present value in the present time $\left<\Lambda(t_{\rm present})\right> \sim E^2_{\rm Planck}/t^2_{\rm present} \sim 10^{-120} E^4_{\rm Planck}$ \cite{KlinkhamerVolovik2008b}, see Fig. 20.
%\ref{VacDecay}.  
This process looks similar to the Starobinsky inflation, except for the magnitude of the oscillation frequency, which in our case is of the Planck scale instead of the Higgs inflaton mass. 

The similar oscillating decay takes place in supercoductors after quench \cite{VolkovKogan1974,Barankov2004,Gurarie2014}, see Fig. 20
%\ref{VacDecay} 
 ({\it bottom}). Such oscillations with the frequency equal the mass (gap) of the Higgs amplitude mode, $\omega=2\Delta$, have been observed \cite{Matsunaga2013,Matsunaga2014}. In superfluids and superconductors the role of the vacuum energy is played by $(\Delta^2(t)- \Delta_0^2)^2$, see Sec. 7.3.6 in \cite{Volovik2003}. Then one has 
$\Lambda(t)\propto \omega^3 \frac {\sin^2\omega t}{t}$, and $\left<\Lambda(t)\right>\propto  \omega^3/t$.

But if the initial conditions are different, then from our equations (again still without dissipation) it follows that the Universe relaxes to the de Sitter spacetime instead of the Minkowski vacuum state.
The question arises: what is the fate of the de Sitter vacuum? Does Hawking radiation from the de Sitter cosmological horizon exist? If the de Sitter vacuum radiates, does the Hawking radiation lead to the decrease of the vacuum energy? Is the de Sitter vacuum unstable? The instability of the de Sitter  vacuum is supported by Polyakov \cite{Polyakov2008,Polyakov2010,KrotovPolyakov2011,Polyakov2018}, but is not supported by Starobinsky.
My view on that problem is in the papers \cite{Volovik2009a}, which is closer to the Starobinsky view.

\section{Conclusion}

The overwhelming majority of my work emerges from the Landau Institute environment, and/or in collaboration with the experimental ROTA group  in the Low Temperature Laboratory of Aalto University.
I did not mention here the inspiration from or/and the direct collaboration with Edel'stein \cite{VolovikMelnikovEdestein1973}, Eliashberg, Kats \cite{KatsVolovik1984}, Khmel'nitskii \cite{VolovikKhmelnitskii1984},  Makhlin \cite{MakhlinVolovik1995,Makhlin1996},
Mel'nikov \cite{VolovikMelnikovEdestein1973}, Pokrovsky, Rashba, Sinai  and the other members of Landau Institute in different areas of physics, who were also gathered by Khalatnikov into the unique Institute.

\section*{Acknowledgements}
This work has been supported by the European Research Council (ERC) under the European Union's Horizon 2020 research and innovation programme (Grant Agreement No. 694248).

\end{document}